\documentclass[runningheads]{llncs}

\usepackage{eccv}

\usepackage{eccvabbrv}

\usepackage{graphicx} \usepackage{booktabs}

\usepackage[accsupp]{axessibility} %

\usepackage[dvipsnames]{xcolor}

\DeclareMathOperator{\diag}{diag}
\DeclareMathOperator{\loss}{\mathcal{L}}

\usepackage[OT1]{fontenc} 

\graphicspath{{./figs/}}

\usepackage{times}
\usepackage{epsfig}
\usepackage{graphicx}
\usepackage{amsmath}

\usepackage{amssymb}
\usepackage{booktabs}
\usepackage{bm}
\usepackage{bbm}
\usepackage{setspace}
\usepackage{caption}
\usepackage{multirow}
\usepackage{extarrows}
\usepackage[normalem]{ulem}
\usepackage{tikz}
\usepackage{xr}
\usepackage{colortbl}
\usepackage{xcolor}

\newcommand{\real}{\mathbb{R}}

\newcommand{\best}[1]{$\textbf{#1}$}
\newcommand{\sbest}[1]{$\underline{\textrm{#1}}$}

\usepackage[pagebackref,breaklinks,colorlinks,citecolor=eccvblue]{hyperref}

\usepackage{tcolorbox}

\usepackage{orcidlink}

\begin{document}

\title{A Compact Dynamic 3D Gaussian Representation \\for Real-Time Dynamic View Synthesis}
\titlerunning{A Compact Dynamic 3D Gaussian}

\author{Kai Katsumata \and Duc Minh Vo \and Hideki Nakayama}

\authorrunning{K.~Katsumata et al.}

\institute{The University of Tokyo, Japan \\
{\tt\small \{katsumata,vmduc,nakayama\}@nlab.ci.i.u-tokyo.ac.jp}}

\maketitle

\begin{abstract}

3D Gaussian Splatting (3DGS) has shown remarkable success in synthesizing novel views given multiple views of a static scene.
Yet, 3DGS faces challenges when applied to dynamic scenes because 3D Gaussian parameters need to be updated per timestep, requiring a large amount of memory and at least a dozen observations per timestep.
To address these limitations, we present a compact dynamic 3D Gaussian representation that models positions and rotations as functions of time with a few parameter approximations while keeping other properties of 3DGS including scale, color and opacity invariant. Our method can dramatically reduce memory usage and relax a strict multi-view assumption. %
In our experiments on monocular and multi-view scenarios, we show that our method not only matches state-of-the-art methods, often linked with slower rendering speeds, in terms of high rendering quality but also significantly surpasses them by achieving a rendering speed of $118$ frames per second (FPS) at a resolution of 1,352$\times$1,014 on a single GPU.

\end{abstract}

\section{Introduction}
\label{sec:intro}

\begin{figure}[t]
     \centering
  \bgroup 
  \def\arraystretch{0.2} 
  \setlength\tabcolsep{0.2pt}
    \begin{tabular}{cccccc}
    \includegraphics[width=0.166\linewidth]{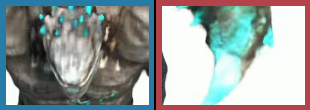} &
    \includegraphics[width=0.166\linewidth]{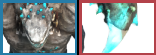} &
    \includegraphics[width=0.166\linewidth]{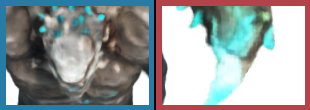} &
    \includegraphics[width=0.166\linewidth]{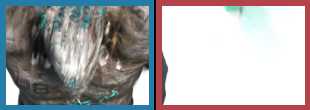} &
    \includegraphics[width=0.166\linewidth]{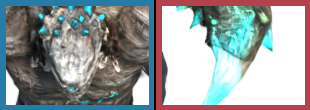} &
    \includegraphics[width=0.166\linewidth]{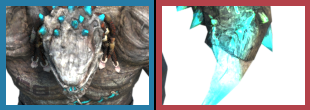} \\
    \includegraphics[width=0.166\linewidth,trim={100px 160px 120px 160px},clip]{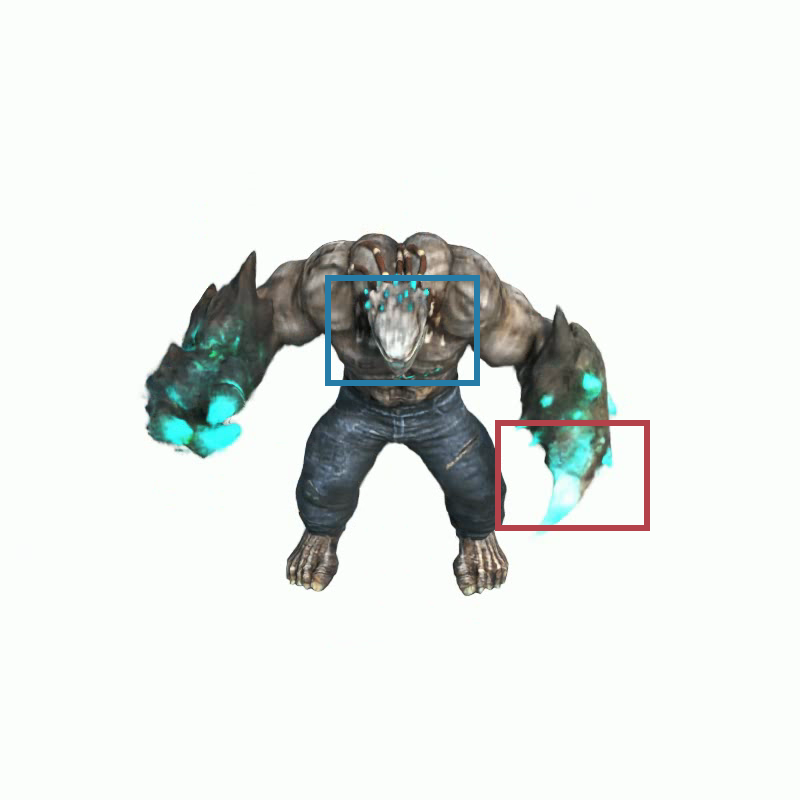} &
    \includegraphics[width=0.166\linewidth,trim={50px 80px 60px 80px},clip]{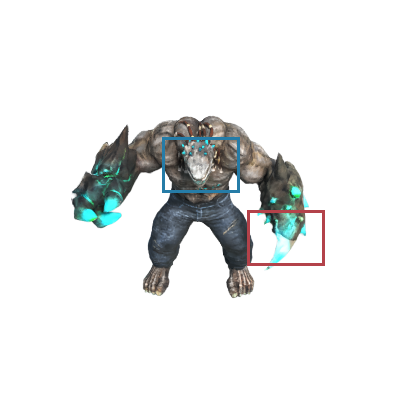} &
    \includegraphics[width=0.166\linewidth,trim={100px 160px 120px 160px},clip]{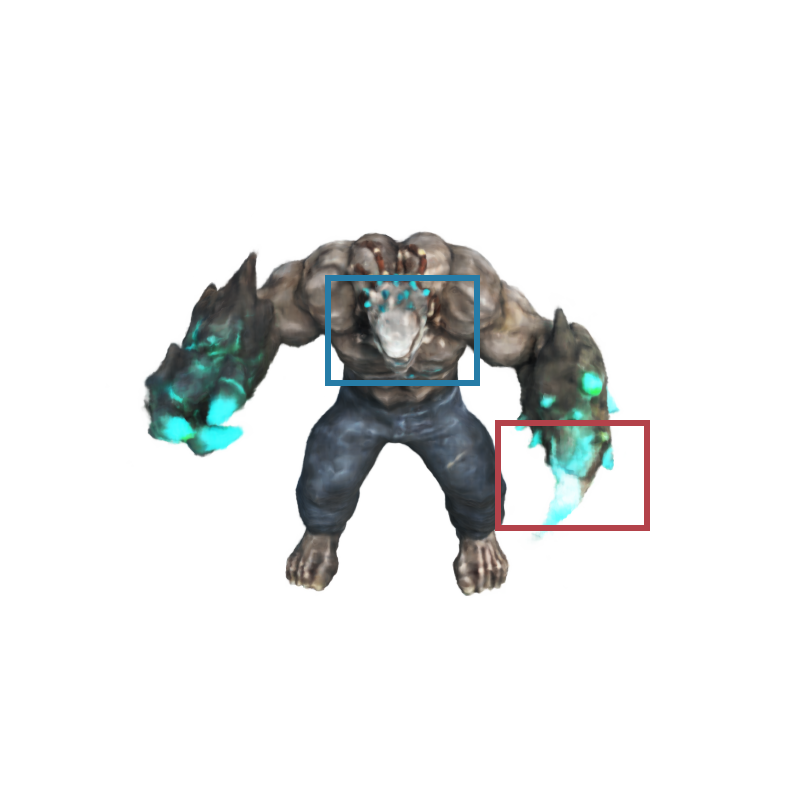} &
    \includegraphics[width=0.166\linewidth,trim={100px 160px 120px 160px},clip]{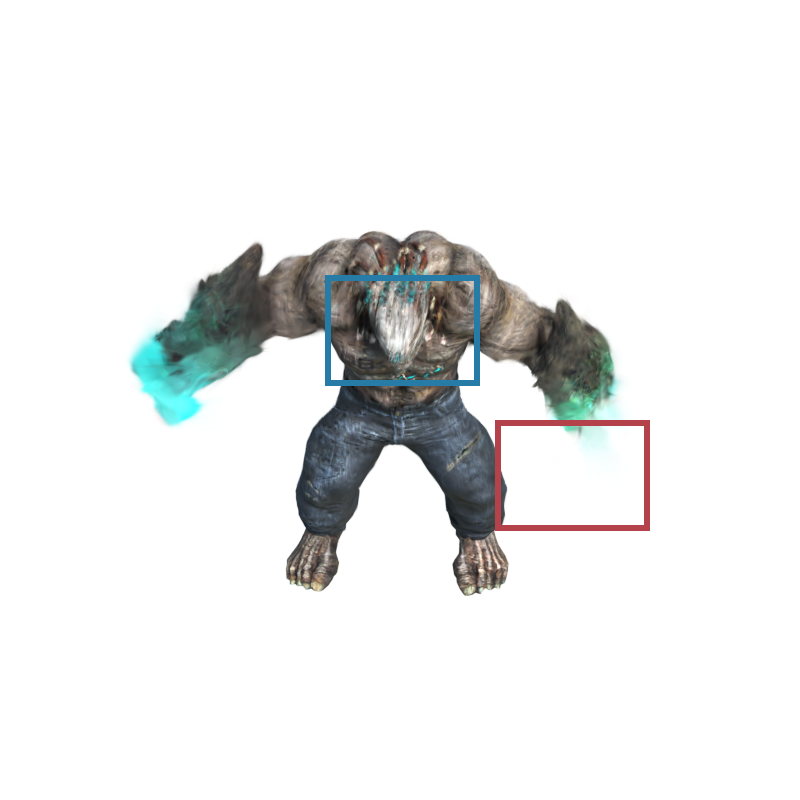} &
    \includegraphics[width=0.166\linewidth,trim={100px 160px 120px 160px},clip]{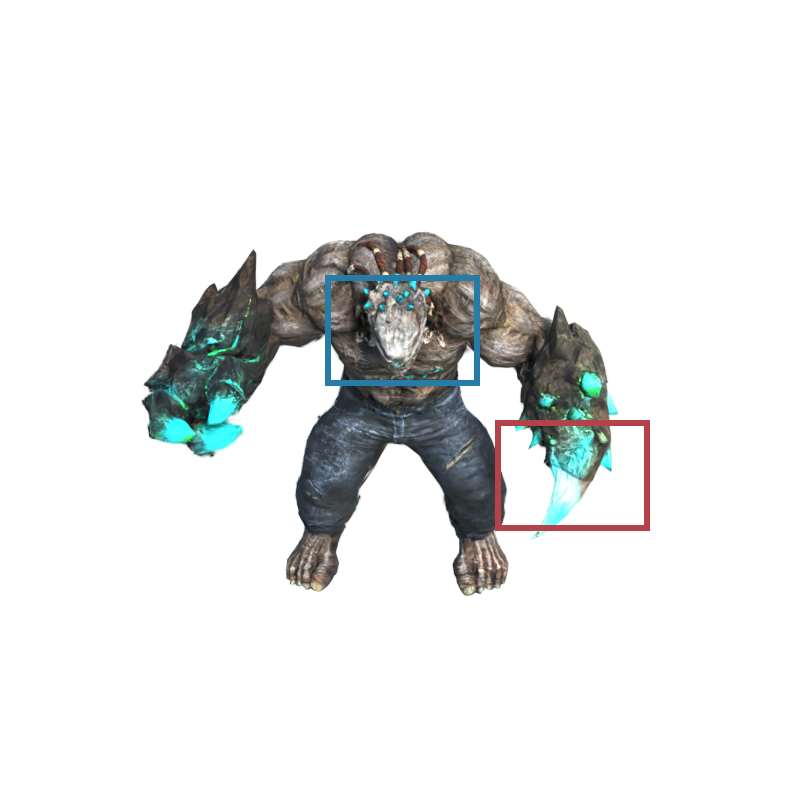} &
    \includegraphics[width=0.166\linewidth,trim={100px 160px 120px 160px},clip]{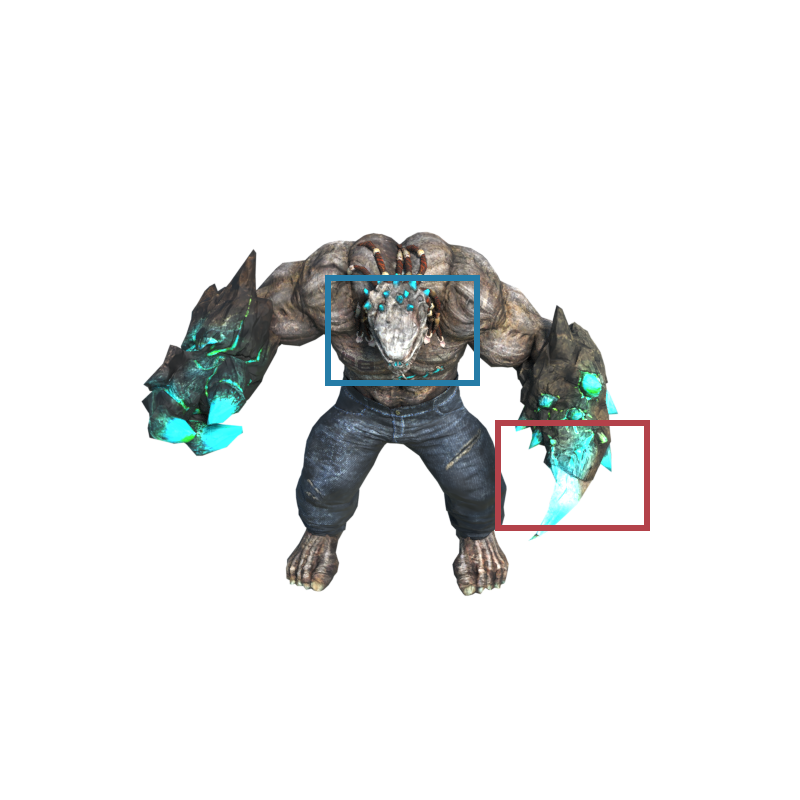}\\
    \small PSNR: 33.56 & \small PSNR: \sbest{36.27} & \small PSNR: 33.61 & \small PSNR: 21.52 & \small PSNR: \best{37.45} & \\\\
    \small FPS: 0.54 & \small FPS: \sbest{1.23} & \small FPS: 0.25 & \small FPS: 169 & \small FPS: \best{146} & \\\\
   \small  Mem: 471MB &\small  Mem: 1.2GB &\small  Mem: \best{48MB} &\small  Mem: 21MB &\small  Mem: \sbest{89MB} & \\\\
        \small K-Planes~\cite{fridovich2023k} & \small V4D~\cite{gan2023v4d} & \small\!\!TiNeuVox-B~\cite{TiNeuVox}\!\! & \small 3DGS~\cite{kerbl20233d} & \small Ours & \small Ground Truth %
  \end{tabular}\egroup
\caption{%
We show examples of novel view synthesis on the \textsc{Mutant} scene in the D-NeRF dataset, visual quality (PSNR), rendering speed (FPS), and memory used to store optimized parameters. 
Our method yields reconstruction fidelity competitive with SoTAs with real-time rendering, achieving $100\times$ faster than V4D and reasonable memory size. 
Non-obvious differences in quality are highlighted. \best{Bold} typeface number indicates the best result among the methods with the competitive rendering quality (excepting for 3DGS), and the \sbest{underline} one does the second best.}\label{fig:teaser} 
\vspace*{-1.5\baselineskip}
\end{figure}

The landscape of novel view synthesis of scenes captured through multiple images/videos has undergone a revolutionary transformation, owing principally to major breakthroughs in neural radiance field (NeRF) approaches~\cite{Mildenhall20eccv_nerf,barron2021mip,tewari2022advances}.
While achieving remarkable visual quality, particularly in dynamic scenes~\cite{pumarola2021d,li2022neural,gao2021dynamic,li2023dynibar,attal2023hyperreel}, NeRFs inevitably confront hurdles in terms of high-speed training and rendering~\cite{Mildenhall20eccv_nerf,reiser2021kilonerf,park2021hypernerf,park2021nerfies}.
This limitation is attributed to the reliance on 
multi-layer perceptrons (MLPs).
Recently, 3D Gaussian Splatting (3DGS)~\cite{kerbl20233d} introduces a differentiable 3D Gaussian representation and point-based rasterization, signaling a departure from neural network reliance.
3DGS emerges as a promising solution that not only accelerates training and rendering processes but also delivers high-quality rendered scenes,
rivaling the levels set by NeRF~\cite{Mildenhall20eccv_nerf} on static scenes.

Nonetheless, in the realm of dynamic scene synthesis, 3DGS encounters a challenge in memory usage and the need for many observations~\cite{luiten2023dynamic}. 
In particular, numerous 3D Gaussian parameters are necessarily stored per timestep, resulting in a non-negligible increase in memory usage and the need for numerous observations per timestep.
This implies the failure in monocular or few-view setups. 
Their strict multi-view assumption requires an advanced facility or experts, lacking the flexibility of capturing setup. 
Exploring 3DGS without multi-view assumption enables dynamic view synthesis with a simple and easy camera setup, which is the primary goal of this study.

To achieve memory-efficient real-time dynamic view synthesis from monocular and multi-view videos, we present a compact dynamic 3D Gaussian representation, 
containing time-invariant and time-varying parameters to capture dynamic motion effectively.
Similarly to~\cite{kerbl20233d,luiten2023dynamic}, we use scaling factors in the covariance matrix, opacity, and color as time-invariant parameters.
Since modeling the change of positions over time is important to represent dynamic scenes~\cite{pumarola2021d,park2021nerfies,park2021hypernerf},
as time-varying parameters, we express each 3D Gaussian position as a function of time to model the temporal change of the position. We also represent 3D Gaussian rotation as a time-varying parameter because the rotation of the objects in the world can be typically changed.
Inspired by the studies that model motion as periodic~\cite{akhter2008nonrigid,zheng2015sparse}, we fit the position by using the Fourier approximation.
We fit the rotation using the linear approximation.
The time-varying parameters make our representation dynamic, meaning that 3D Gaussian moves and rotates over time. 
Moreover, as we use a function with a few parameters to represent the position, the small degree of freedom contributes to the smoothness of reconstructed scenes, enhancing the robustness against unseen views.
Importantly, our representation's memory consumption is solely determined by the number of 3D Gaussians and the number of the approximation function's parameters, independent of the input length. 
Beyond optimizing Gaussian representations through image-level reconstruction, we further enhance temporal consistency by supervising the Gaussian with optical flow obtained from input videos. 
This ensures high-quality reconstruction and facilitates the generalization of the representation.

Our experiments on dynamic datasets, including D-NeRF~\cite{pumarola2021d}, DyNeRF~\cite{li2022neural}, and HyperNeRF~\cite{park2021hypernerf} datasets, demonstrate the effectiveness of optimizing our dynamic 3D Gaussian from both monocular and multi-view videos, showing that our proposed method achieves rendering quality that rivals previous NeRFs~\cite{fridovich2023k,TiNeuVox,gan2023v4d}.
In addition to faithful rendering quality, the proposed method achieves rendering speeds similar to a fast radiance field method~\cite{kerbl20233d} while avoiding large memory increases caused by a dynamic extension (see \cref{fig:teaser}).
Finally, we show an editing application enabled by the explicit property of 3D Gaussian representations.
To sum up, our contributions are:
\begin{itemize}
    \item We present a compact dynamic 3D Gaussian representation with time-varying Gaussian parameters equipped with basis functions for representing dynamic scenes.
    \item Since 3D Gaussian representation is defined over all the timesteps, the 3D Gaussian parameters can be optimized with the frames at all the timesteps, enabling dynamic scene reconstruction from monocular or few-view videos.
    \item Our dynamic 3D Gaussian representation facilitates real-time high-quality dynamic scene rendering of high-resolution images of 1,352$\times$1,014 with a frame rate of 118 FPS using a single GPU. 
    
\end{itemize}

\section{Related Work}
We first briefly overview the radiance field for dynamic
scenes, then discuss the recent efforts for efficient radiance fields with explicit representation, such as grid-, plane-, hash-, and point-based methods, placing our work in real-time dynamic view synthesis.

\subsection{Dynamic view synthesis}
Applications in virtual reality and computer vision often need reconstruction
of dynamic scenes.
Several works extend NeRF~\cite{Mildenhall20eccv_nerf} to handle dynamic scenes in multi-view or monocular setups by time-varying NeRF~\cite{pumarola2021d,li2022neural,gao2021dynamic,tretschk2021non}.
The regularization techniques for temporal smoothness enable suitable scene representations from monocular 
videos~\cite{li2021neural}.
Additional sensory information is also useful for spatio-temporal regularization.
Some attempts~\cite{tian2023mononerf,li2021neural,gao2021dynamic} employ depth or flow, which are observed or predicted with external networks to reconstruct the scene from sparse observations.
Deformation-based approaches~\cite{tretschk2021nrnerf,park2021nerfies,park2021hypernerf,song2023nerfplayer}, another direction of the efforts for dynamic reconstruction, combine static NeRF and deformation fields. 
Although tremendous efforts show high visual quality for dynamic view synthesis, the need for many querying to MLP of NeRFs results in the drawback of slow optimization and rendering~\cite{yiheng2022neural}. 
Our study aims to enable real-time dynamic view synthesis with high visual quality.
We aim to extend 3DGS to dynamic scene reconstruction to achieve high-speed rendering while maintaining the rendering quality from sparse training views. 

\subsection{Explicit Radiance Fields}

A recent line of work~\cite{yu2021plenoctrees,fridovich2022plenoxels,chen2022tensorf} addresses the issue in implicit models (\ie, NeRFs) by exploring explicit models, 
reducing optimization and rendering time.
Plenoxels~\cite{fridovich2022plenoxels} directly optimizes 3D grid representation instead of neural networks. Generally, explicit models 
sacrifice visual quality for fast training~\cite{fridovich2022plenoxels}. 
Hybrid approaches~\cite{chen2022tensorf,muller2022instant,fridovich2023k,shao2023tensor4d,TiNeuVox,gan2023v4d} aim to achieve better trade-offs between training time and visual quality. 
Instant-NGP allows a compact MLP by exploiting a multi-level hash grid to encode positions to feature vectors~\cite{muller2022instant}.
Plane-based approaches are principally designed for representing bounded scenes~\cite{chan2022efficient,an2023panohead,he2023orthoplanes,dong2023ag3d,fridovich2023k,cao2023hexplane}. MERF~\cite{reiser2023merf} employs a multiresolution representation and a fast contraction function to reconstruct unbounded scenes.
For dynamic scenes, K-planes~\cite{fridovich2023k} decomposes 4D dynamic volumes into multiple feature planes and employs an MLP-based feature decoder for determining color and density. 
The structured representation hinders rendering speed by many querying to the representation.
In this study, unstructured 3D Gaussians promise large gains in rendering speed.

\subsection{Point-based rendering}
Points, which naturally come from depth sensors, Structure from Motion (SfM)~\cite{schoenberger2016sfm}, or common Multi-View Stereo (MVS) algorithms~\cite{schonberger2016pixelwise,seitz2006comparison}, offer a useful representation for representing fine-grained scenes and complex objects, as well as facilitating computationally efficient rendering. Due to these aspects of the representation, they have been well-studied in the vision and graphics community.
The differentiable pipeline for point-based rendering results in that points can be used for reconstructing 3D scenes~\cite{keselman2022approximate,kopanas2021point,kerbl20233d,kopanas2022neural}. 
3DGS~\cite{kerbl20233d} achieves real-time rendering with high visual quality for unbounded static scenes at the expense of generalization performance derived from NeRF's continuous neural field representation.
3DGS is replacing NeRFs as the backbone of text-to-3D models, leading to faster 3D generation~\cite{tang2023dreamgaussian,chen2023text,yi2023gaussiandreamer}.
Recently, Dynamic 3D Gaussians~\cite{luiten2023dynamic} employs 3DGS for dynamic scenes, which models dynamic scenes by the Gaussian position and rotation at each timestamp. 
The position and rotation of Gaussians at every timestamp are
effective in modeling scenes from dense multi-view dynamic scenes. However,
this approach presents difficulties in reconstructing monocular dynamic
scenes, resulting in excessive memory consumption, particularly for
extended input sequences. %
Specifically, the space complexity of the method for the scene with $T$ frames is $O(TN)$, where N is the number of 3D Gaussians. Our goal is to reduce memory consumption by representing time-varying position and rotation with approximation using a few parameters.
The space complexity of our method is $O(LN)$, where $L$ is the number of parameters of the approximation, and usually $L < T$.

Concurrent with this work, SpacetimeGaussian~\cite{li2023spacetime} addresses dynamic view synthesis from mulit-view, unlike this study, videos by combining Gaussian Splatting and MLPs. \cite{wu20234d} aims to model motion by employing a deformation field network while sacrificing rendering speed. \cite{yang2023real} splits Gaussians in a time direction, and each Gaussian only focuses a local temporal space.
4D-Rotor Gaussian Splatting~\cite{duan20244d} models a local temporal space via temporal slicing for fast rendering.
We aim to build a memory-efficient Gaussian representation for dynamic scenes, even for monocular scenes, while keeping pure 3D Gaussian representation in order not to sacrifice the gift of 3D Gaussians, such as outstanding rendering speed and ease of direct editing of the scene. 

\section{Method}

\begin{figure}[t]
  \centering
    \includegraphics[width=0.9\linewidth]{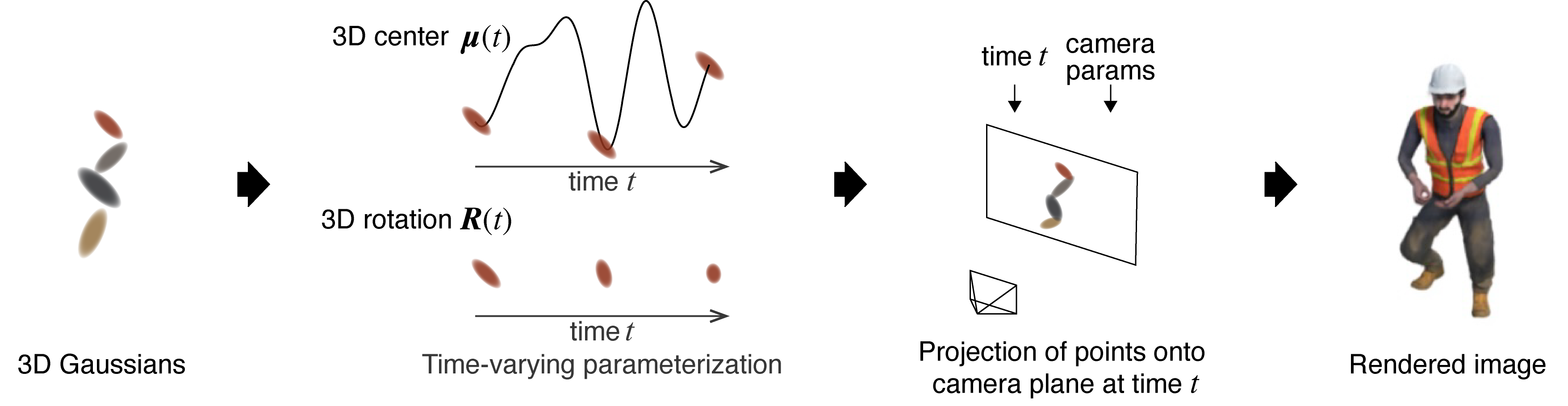}
\caption{Overview of our dynamic view synthesis framework. Our dynamic 3D Gaussian representation takes temporal modeling of 3D centers and rotations with Fourier and Linear approximation, respectively. Our representation parameters are shared over all the timesteps, and observations of each timestep hint at the representation for other timesteps, enabling compact representation and reconstruction of dynamic scenes from few-view videos. 
In this figure, we only illustrate the time-varying parameterization of one Gaussian for the sake of simplicity.
 }\label{fig:method} 

\end{figure}

Given images with timesteps and camera parameters obtained from videos, our task aims to learn a 4D spatial-temporal representation of a dynamic scene, which enables fast and high-quality view rendering.
To achieve this, we make 3DGS meet dynamic view synthesis.
The original 3D Gaussian representation~\cite{kerbl20233d} is defined by 
a position (mean), a covariance matrix (decomposed into a rotation matrix and a scaling vector)%
, a color (determined by spherical harmonics (SH)~\cite{brian1987} coefficient), and an opacity.
To represent dynamic scenes,
each 3D Gaussian in our method (\cref{fig:method}) regards the position and rotation as time-varying parameters and others as time-invariant parameters over
time (\cref{sec:representation}).
Given the Gaussians, intrinsic and extrinsic camera parameters, and a timestep, we render images with the 3DGS technique~\cite{kerbl20233d},
which renders an image by employing Gaussians within the camera plane out of a set of Gaussians (\cref{sec:rendering}).
We update the Gaussian parameters to decrease the distance between rendered and training images in image and flow spaces (\cref{sec:optimization}). Flow reconstruction loss enhances the temporal consistency of the learned representation, resulting in plausible image reconstruction.
The small degrees of freedom of our representation essentially facilitate the reconstruction of dynamic scenes from a few observations.

\subsection{Dynamic 3D Gaussian representation}\label{sec:representation}
One straightforward extension of 3DGS ~\cite{luiten2023dynamic}
to dynamic scenes is to model the scenes per timestep explicitly. Although that strategy enables flexible modeling for dynamic scenes, it needs 3D Gaussian parameters per timestep, increasing the memory size in proportion to video length. Since the representation for each time is optimized by observations with the number of cameras, the strategy does not get a sufficient number of observations in the case of monocular or few-view videos, implying the failure in monocular or few-view camera setup.

To design a compact dynamic 3D Gaussian representation, we aim to express 3D Gaussian parameters using only a few parameters to achieve faithful reconstruction without a large increase in parameters.
Our dynamic scene representation comprises a set of dynamic 3D Gaussians, extending the static 3D Gaussian introduced in~\cite{kerbl20233d}.
Our dynamic representation lets the 3D Gaussians move through the scene over time, using time-varying parameters (center position and rotation factors) and time-invariant parameters (scale, color, and opacity). 
Each dynamic Gaussian 
encapsulates:

\noindent
\begin{minipage}{\columnwidth}
\centering
\vspace{0.5mm}
\begin{tcolorbox}[top=0mm,bottom=0mm,left=2mm,right=2mm]
\raggedright
1) a 3D center at time $t$: $[x(t), y(t), z(t)]^{\sf T} \in \real^{3}$, \\
2) a 3D rotation at time $t$ represented by a quaternion: $[q_x(t), q_y(t), q_z(t), q_w(t)]^{\sf T} \in \real^{4}$, \\ 
3) a scaling factor: $[s_x, s_y, s_z]^{\sf T} \in \real^{3}$,\\
4) spherical harmonics coefficients representing color with the degrees of freedom $k$: $h \in \real^{3 \times (k+1)^2}$, \\
5) an opacity: $o \in \real$.
\end{tcolorbox}
\vspace{0.5mm}
\end{minipage}

Each Gaussian at time $t$ is characterized by a 3D center $\bm{\mu}(t) = \left[x(t),  y(t), z(t) \right]^{\sf T}$ and a 3D covariance matrix $\mathbf{\Sigma}(t)$. The density of 3D Gaussian at the intersection $\vec{\bm{x}}$ with a ray is obtained
as follows:
\begin{align}
  G_{t}(\vec{\bm{x}}) = e^{-\frac{1}{2}(\vec{\bm{x}} - \bm{\mu}(t))^{\sf T} \mathbf{\Sigma}(t)^{-1} (\vec{\bm{x}}-\bm{\mu}(t))}.
\end{align}
To constrain the covariance matrix $\mathbf{\Sigma}(t)$ to be a positive semi-definite matrix during optimization,
the covariance matrix $\mathbf{\Sigma}(t)$ is decomposed by using a scaling matrix $\mathbf{S} = \diag(s_x, s_y, s_z)$  and a rotation matrix $\mathbf{R}(t)$ as $\mathbf{\Sigma}(t) = \mathbf{R}(t)\mathbf{S}\mathbf{S}^{\sf T}\mathbf{R}(t)^{\sf T}$.
Here, a rotation matrix $\mathbf{R}(t)$ is represented by quaternion $(q_x(t), q_y(t), q_z(t), q_w(t))$.
Since most parts of the dynamic scene hardly change in scale because the solid (\eg, humans, animals, and things) scarcely expands or shrinks, we maintain the scale parameter as a constant to reduce the model size.
In what follows, we formally define the 3D center and rotation.

Since motion in dynamic scenes is primarily described by changing the position of points like scene or optical flow~\cite{vedula1999three,lucas1981iterative}, we model 3D center with expressive approximation.
We approximate the 3D position $x(t), y(t), z(t)$ using Fourier approximation.
At time $t$, it is represented by
\begin{equation}
\begin{split}
  x(t) & = w_{x,0} + \sum_{i=1}^{L} w_{x,2i-1}\sin (2 i \pi t) + w_{x,2i} \cos(2 i \pi t), \\
  y(t) & = w_{y,0} + \sum_{i=1}^{L} w_{y,2i-1}\sin (2 i \pi t) + w_{y,2i} \cos(2 i \pi t), \\
  z(t) & = w_{z,0} + \sum_{i=1}^{L} w_{z,2i-1}\sin (2 i \pi t) + w_{z,2i} \cos(2 i \pi t), \label{eq:center}
\end{split}
\end{equation}
where, $w_{\cdot,0},\ldots,w_{\cdot,2L}$ are intercept and coefficients of the position, and $L$ is the number of terms (harmonics). 
We remark that a polynomial approximation is inadequate due to underfitting with a small number of bases and overfitting with higher-order polynomials.
This aspect makes us choose the Fourier approximation.

3DGS uses anisotropic 3D Gaussians, resulting in the need for dynamic modeling of Gaussian rotations.
We approximate the 3D rotation (quaternion) over time using a linear approximation because defining 3D rotation with a unit quaternion makes applying complex approximation difficult.
At time $t$, it is defined as
\begin{equation}
 \begin{split}
  q_x(t) &= w_{qx,0} + w_{qx,1} t, \quad  q_y(t) = w_{qy,0} + w_{qy,1} t,\\
  q_z(t) &= w_{qz,0} + w_{qz,1} t, \quad  q_w(t) = w_{qw,0} + w_{qw,1} t,
   \end{split}
\end{equation}
where $w_{\cdot,0}$ and $w_{\cdot,1}$ are intercepts and coefficients of the rotation, respectively.

For each Gaussian, the preceding definitions yield $3L + 8 + 3 + 3(k+1)^2 + 1$ parameters with respect to 3D center, 3D rotation, scale, color, and opacity.
Notably, the parameter count for each Gaussian is defined merely by the number of approximation terms and spherical harmonic degrees of freedom, with no regard to time length.
Compared to storing parameters for each timestep, our approach saves on memory usage.
Memory consumption in our dynamic scene representation is determined by two hyperparameters (\ie, $L$ and $k$) and the number of Gaussians used.
Furthermore, the representation defined as a function of time over continuous time inhibits discontinuous movement through time.
This characteristic improves robustness in novel view synthesis settings.

\subsection{Rendering via 3D Gaussian Splatting}\label{sec:rendering} 

Rendering with 3D Gaussian applies splatting techniques~\cite{kerbl20233d} to the Gaussian within the camera planes.
Zwicker \etal~\cite{zwicker2002ewa} give the projection of the 3D covariance matrix to the 2D one.
The 3D covariance matrix $\mathbf{\Sigma}$ is projected into 2D one $\mathbf{\Sigma}'$ given a viewing transformation $\mathbf{W}$ as $\mathbf{\Sigma}'(t) = \mathbf{J}\mathbf{W}\mathbf{\Sigma}(t) \mathbf{W}^{\sf T} \mathbf{J}^{\sf T}$, 
where $\mathbf{J}$ is the Jacobian of the affine approximation of the projective transformation at Gaussian center $\bm{\mu}(t)$:
\begin{align}
  \mathbf{J} = \begin{bmatrix}
    \frac{1}{v_z} & 0 & - \frac{v_x }{v_z^2} \\
    0 & \frac{1}{v_z} & - \frac{v_y}{v_z^2} \\
    0 & 0 & 0 \\
    \end{bmatrix},\label{eq:jacobian}
\end{align}
where $[v_x, v_y, v_z]^{\sf T} = \mathbf{W}\bm{\mu}(t)$ is the camera coordinate of the Gaussian center $\bm{\mu}(t)$ obtained by the viewing transformation, which projects the points from the world space to the camera space.

Similar to NeRF style volumetric rendering, the point-based rendering computes the color $C$ of a pixel by evaluating the blending of
$N$ ordered points that overlap the pixel $C = \sum_{i=1}^{N} c_i \alpha_i \prod_{j=1}^{i-1} (1 - \alpha_j)$,  
where $c_i$ represents the color of a Gaussian evaluated by SH coefficients, and $\alpha_i$ represents the density that is calculated from
a 2D Gaussian with covariance $\mathbf{\Sigma}'$ at time $t$ and a 2D center $\bm{\mu}'$ at time $t$ and an optimized opacity $o$.

\subsection{Optimization of the dynamic 3D Gaussian representation}\label{sec:optimization}

We optimize the Gaussian parameters: intercepts and coefficients of position and rotation $w$, a scaling factor $s_x, s_y, s_z$, SH coefficients $h$, and an opacity $o$, based on the iterations of rendering and comparing the rendered images with
training frames in the captured videos. To compare the rendered and training views, the loss function contains the L1 loss and  
the Structural Similarity (SSIM)~\cite{wang2004image} loss $\loss_{\rm D-SSIM}$:
\begin{align}
  \loss_{\rm recon} = (1-\lambda) | \hat{I} - I | + \lambda \loss_{\rm D-SSIM},
\end{align}
where $\hat{I}$ and $I$ are rendered and target images, respectively.
The loss function moves and rotates the anisotropic Gaussians and changes their color and opacity so that each Gaussian covers a homogeneous area.
Since the loss just fixes incorrectly positioned Gaussians, the over- or under-representation of the set of Gaussians for the scene needs a mechanism for creating Gaussians that reconstruct the scene or destroy extra Gaussians.
We also follow the divide and prune techniques in 3DGS for producing a compact and precise representation of the scene.
We surveil the gradients of each Gaussian and densify Gaussians by splitting a Gaussian with a large gradient and a large scale into two small Gaussians and cloning a Gaussian with a large gradient and a small scale to two Gaussians. Moreover, we remove transparent Gaussians with an opacity less than a threshold value of 0.005.

Following~\cite{kerbl20233d}, we initialize a set of Gaussians using a set of sparse points from SfM~\cite{schoenberger2016sfm} for real scenes, and we initialize a set of Gaussians randomly using a uniform distribution for synthetic scenes due to the absence of the prior.
We adopt a two-stage optimization strategy consisting of static and dynamic stages.
Deeming the frames in the captured datasets as static scenes, we optimize static representation in the static stage to learn the prior of Gaussians. In other words, we optimize the parameters consistent all over time (\ie, scale, SH coefficients, and opacity) and
intercepts for a center and a rotation ($w_{x,0}, w_{y,0}, w_{z, 0}, w_{qx,0}, w_{qy,0}, w_{qz,0}, w_{qw,0}$)
out of the Gaussian parameters in the static stage.
After the static stage, we optimize all the parameters of the set of Gaussians to reconstruct a dynamic region as a dynamic stage.

Another challenge in the dynamic scene reconstruction is ambiguity caused by the limited number of captured views at a timestep. Since a dynamic scene contains temporal changes, such as moving objects and changing shapes, sharing the scene information over frames with different timesteps is hard.
To overcome the ambiguity, we employ flow information. 
Like our 3D Gaussian, the scene flow~\cite{vedula2005three,mayer2016large,menze2015object} is defined as the position of a point in 3D space and its motion. These 3D points originate from different mechanism than those in 3D Gaussian, making matching in 3D space difficult. Since the optical flow defined on the image plane can be directly matched with 3D Gaussian and is straightforward to compute from monocular inputs, we supervise the flows of the optimizable Gaussians with the ground truth optical flows of the input frames. 
We use the RAFT algorithm~\cite{teed2020raft} to obtain ground truth flow for training views: forward flow $f_{\rm fwd}$ and backward flow $f_{\rm bwd}$ between two adjacent frames. The flow loss $\loss_{\rm flow}$ takes the L1 loss between the ground truth flows and the optical flow of the Gaussian for both directions of the flows.
The flow loss provides the spatial-temporal consistency to our method without any additional computation cost in rendering. %
We combine the flow loss $\loss_{\rm flow}$ with the reconstruction loss that compares the rendered and training views:
\begin{align}
 \loss = \loss_{\rm recon}+ \lambda_{\rm flow}\loss_{\rm flow}(\hat{F}, F),
\end{align}
where, $F = \{f_{\rm fwd}, f_{\rm bwd}\}$ and $\hat{F}$ are the ground truth flow and the flow of the Gaussians, respectively, and $\lambda_{\rm flow}$ is a balancing hyperparameter for the flow term.
Instead of applying an optical flow algorithm for rendering, we create pseudo optical flow from Gaussian representation.
The motion of the scene is represented by only the coefficients of the 3D Gaussian mean
$w_{x,1\leq i}, w_{y,1\leq i}, w_{z,1\leq i}$.
We can calculate the scene flow in the 3D space by
\begin{equation}
 \begin{split}
  \hat{f}^{x}_{\rm fwd} = x(t + \Delta t) - x(t), \quad \hat{f}^{x}_{\rm bwd} = x(t) - x(t - \Delta t), \\
  \hat{f}^{y}_{\rm fwd} = y(t + \Delta t) - y(t), \quad \hat{f}^{y}_{\rm bwd} = y(t) - y(t - \Delta t), \\
  \hat{f}^{z}_{\rm fwd} = z(t + \Delta t) - z(t), \quad \hat{f}^{z}_{\rm bwd} = z(t) - z(t - \Delta t), 
\end{split}
\end{equation}
where $\Delta t$ is the difference between the timesteps of the two image frames.
The scene flow is projected into a 2D camera plane using
\begin{align}
  \hat{f}^{\rm xyz}_{\{\rm fwd,bwd\}} = \mathbf{J} [\hat{f}^{x}_{\{\rm fwd,bwd\}}, \hat{f}^{y}_{\{\rm fwd,bwd\}}, \hat{f}^{z}_{\{\rm fwd,bwd\}}]^{\sf T},
\end{align}
where $\mathbf{J}$ is the Jacobian of the affine approximation of the projective transformation at the Gaussian center $\bm{\mu}$ (\cref{eq:jacobian}).
Regarding the scene flows on the camera plane as RGB colors, we can apply the point-based rendering to calculate an optical flow of a pixel as the $\alpha$-blending:
\begin{align}
  \hat{f}_{\rm fwd} = \sum_{i=1}^{N} \hat{f}^{\rm xyz}_{{\rm fwd},i} \alpha_i \prod_{j=1}^{i-1} (1 - \alpha_j).
\end{align}
The backward flow is calculated in the same way. The optical flow $\hat{F}$ consists of the forward flows $\hat{f}_{\rm fwd}$ and backward flows $\hat{f}_{\rm bwd}$
for all pixels. 
We exclude the flow loss for the D-NeRF dataset because the teleport of the cameras between neighbouring frames causes difficulties in calculating ground truth flows.

\begin{figure*}[tb]
  \centering
  \bgroup 
  \def\arraystretch{0.0} 
  \setlength\tabcolsep{0.2pt}
  \begin{tabular}{ccccccc}
    &\small Ground Truth &\small K-Planes~\cite{fridovich2023k} &\small V4D~\cite{gan2023v4d} &\small \!\!TiNeuVox-B~\cite{TiNeuVox}\!\! &\small 3DGS~\cite{kerbl20233d} &\small Ours \\
    \raisebox{0.2em}{\smash{\rotatebox[origin=c]{90}{\footnotesize \textsc{T-Rex}}}} & 
    \includegraphics[width=0.160\linewidth,trim={80px 180px 80px 180px},clip]{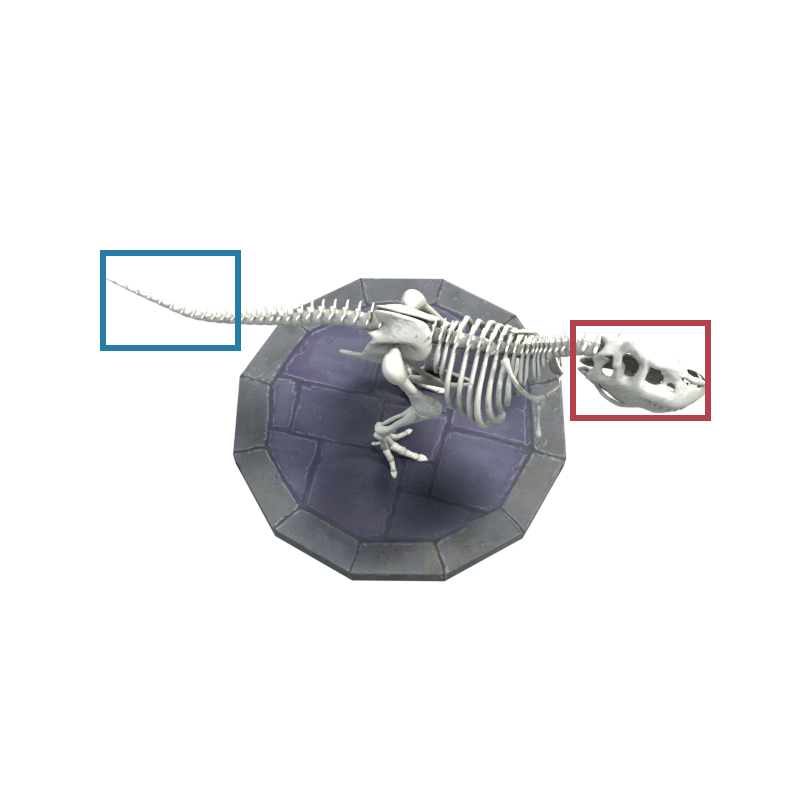} &
    \includegraphics[width=0.160\linewidth,trim={80px 180px 80px 180px},clip]{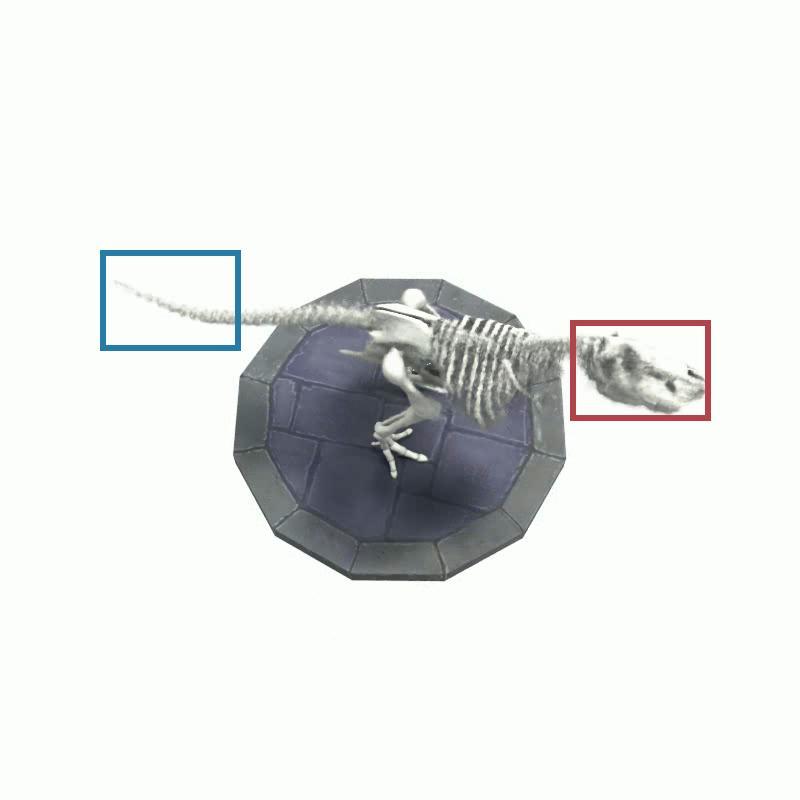} &
    \includegraphics[width=0.160\linewidth,trim={40px 90px 40px 90px},clip]{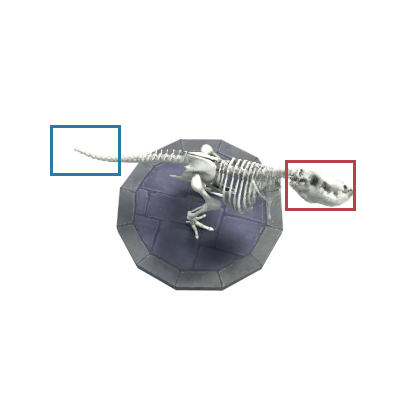} &
    \includegraphics[width=0.160\linewidth,trim={80px 180px 80px 180px},clip]{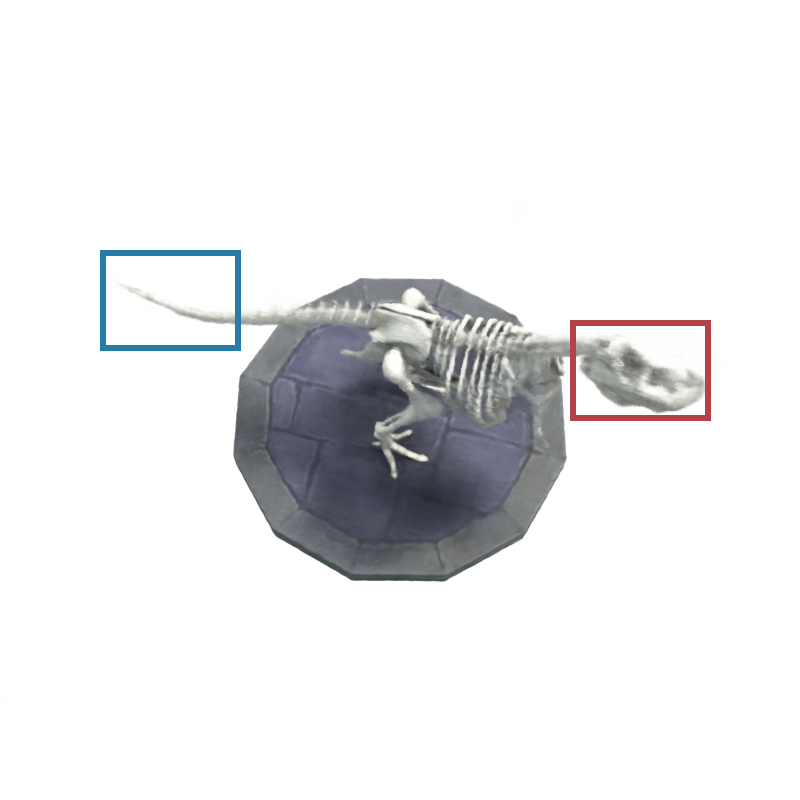} &
    \includegraphics[width=0.160\linewidth,trim={80px 180px 80px 180px},clip]{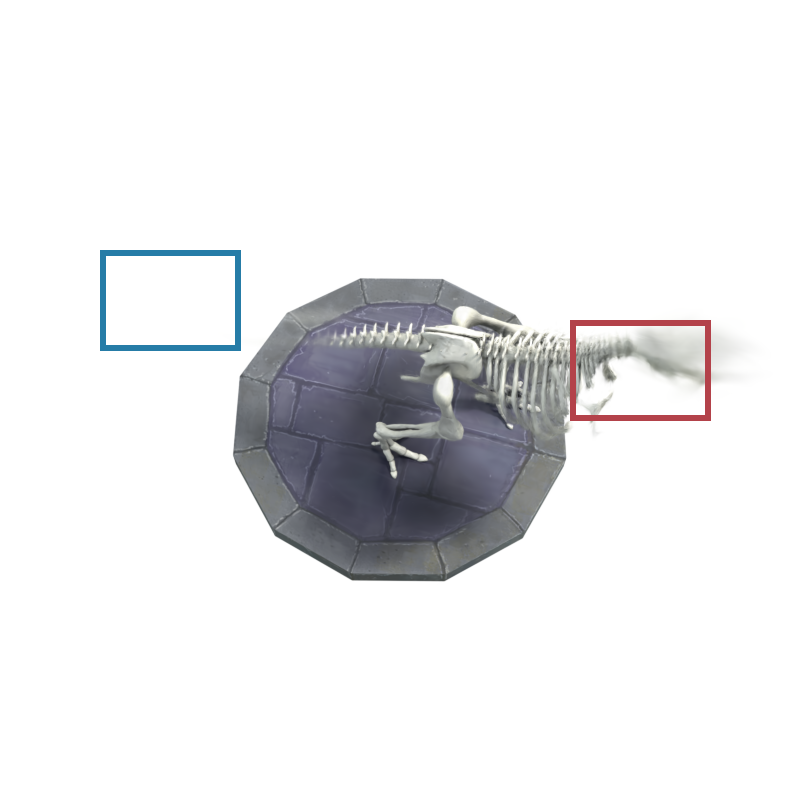} &
    \includegraphics[width=0.160\linewidth,trim={80px 180px 80px 180px},clip]{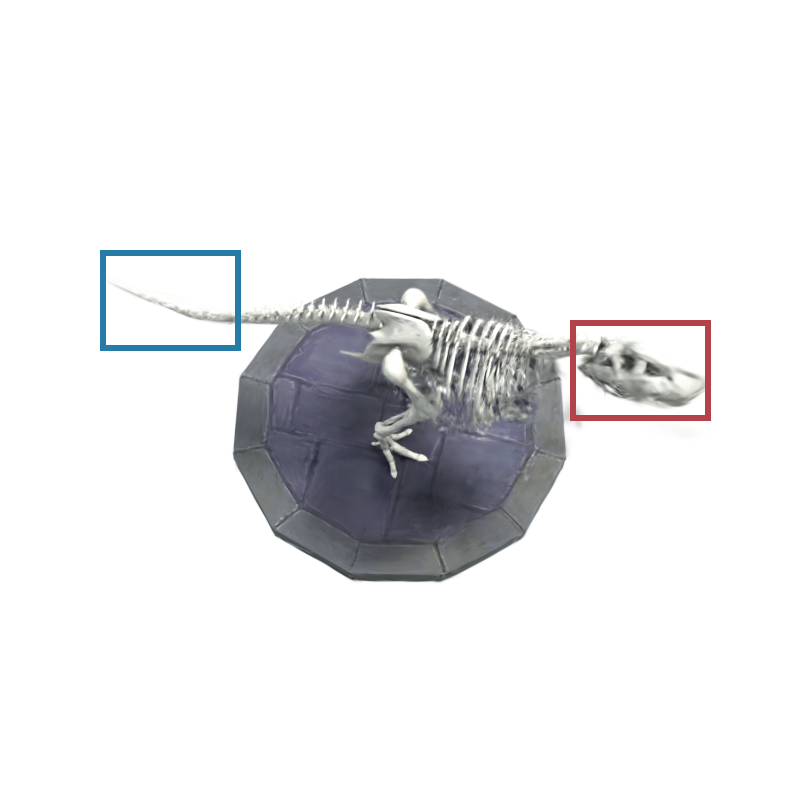} \\ &
    \includegraphics[width=0.160\linewidth]{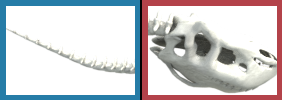} &
    \includegraphics[width=0.160\linewidth]{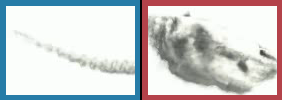} &
    \includegraphics[width=0.160\linewidth]{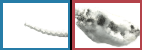} &
    \includegraphics[width=0.160\linewidth]{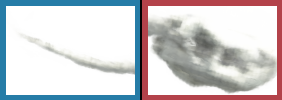} &
    \includegraphics[width=0.160\linewidth]{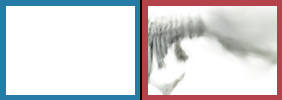} &
    \includegraphics[width=0.160\linewidth]{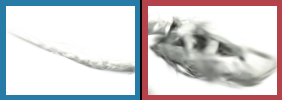} \\
    \raisebox{0.8em}{\smash{\rotatebox[origin=c]{90}{\footnotesize \textsc{Stand Up}}}} &
    \includegraphics[width=0.160\linewidth,trim={200px 220px 200px 280px},clip]{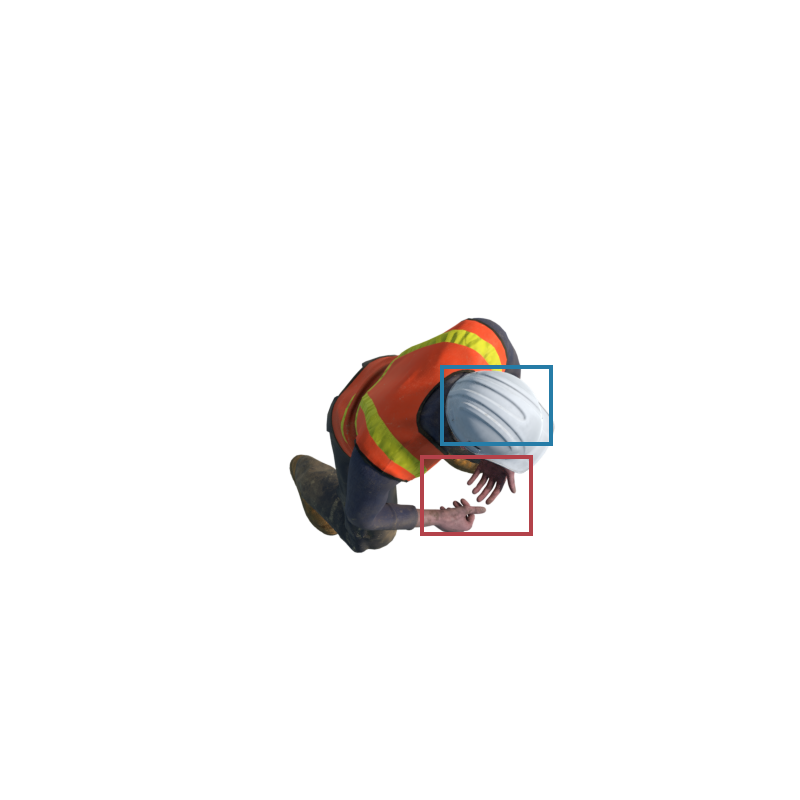} &
    \includegraphics[width=0.160\linewidth,trim={200px 220px 200px 280px},clip]{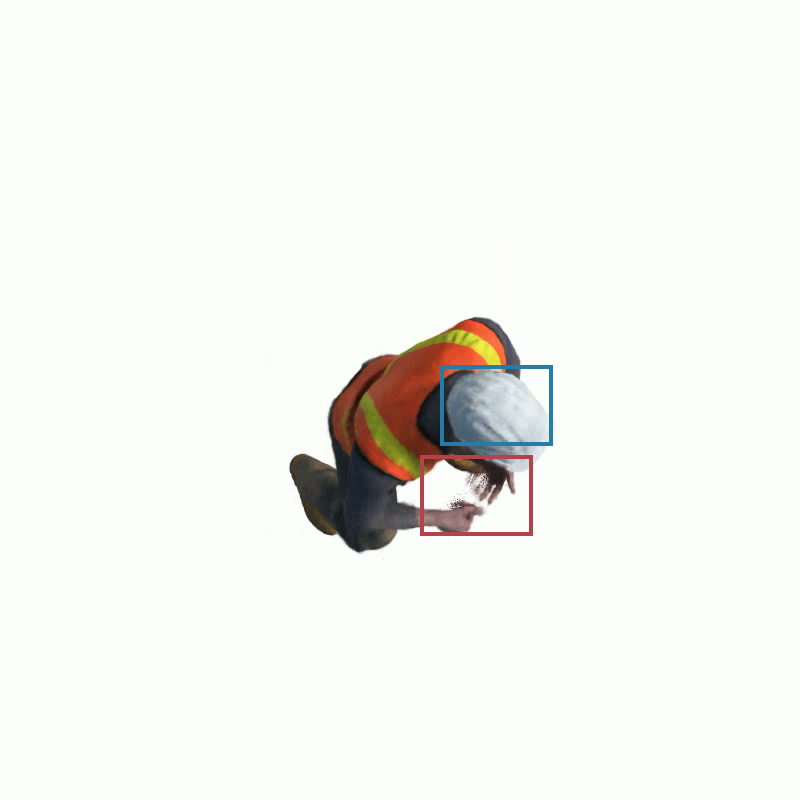} &
    \includegraphics[width=0.160\linewidth,trim={100px 110px 100px 140px},clip]{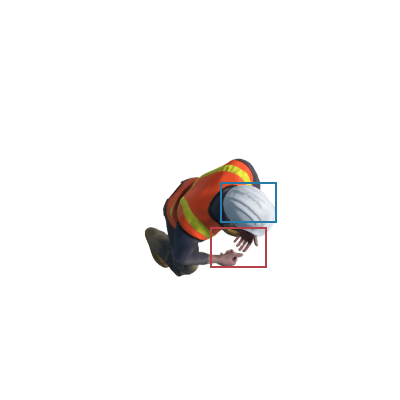} &
    \includegraphics[width=0.160\linewidth,trim={200px 220px 200px 280px},clip]{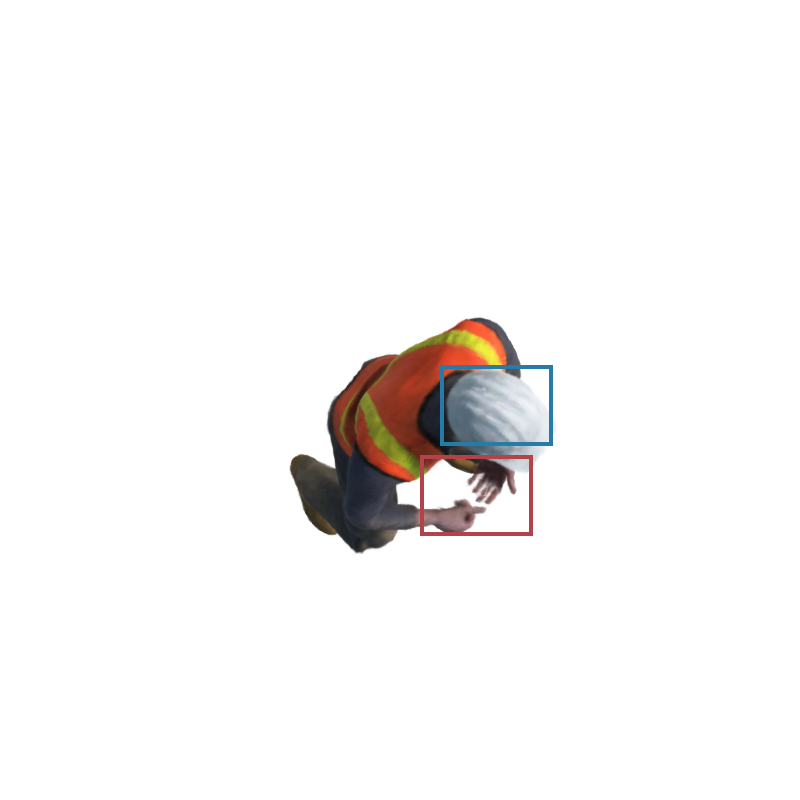} &
    \includegraphics[width=0.160\linewidth,trim={200px 220px 200px 280px},clip]{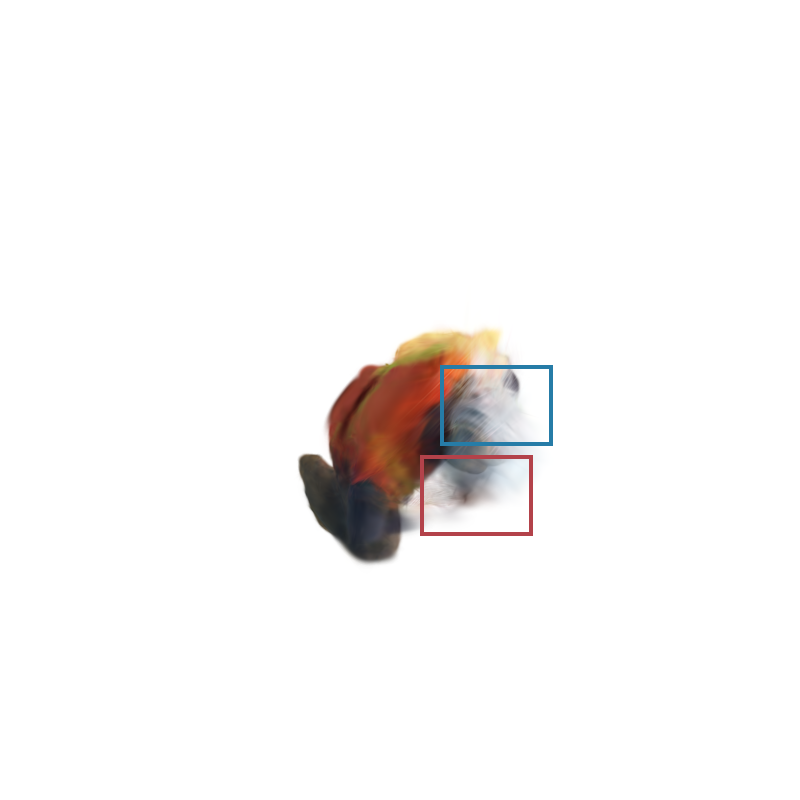} &
    \includegraphics[width=0.160\linewidth,trim={200px 220px 200px 280px},clip]{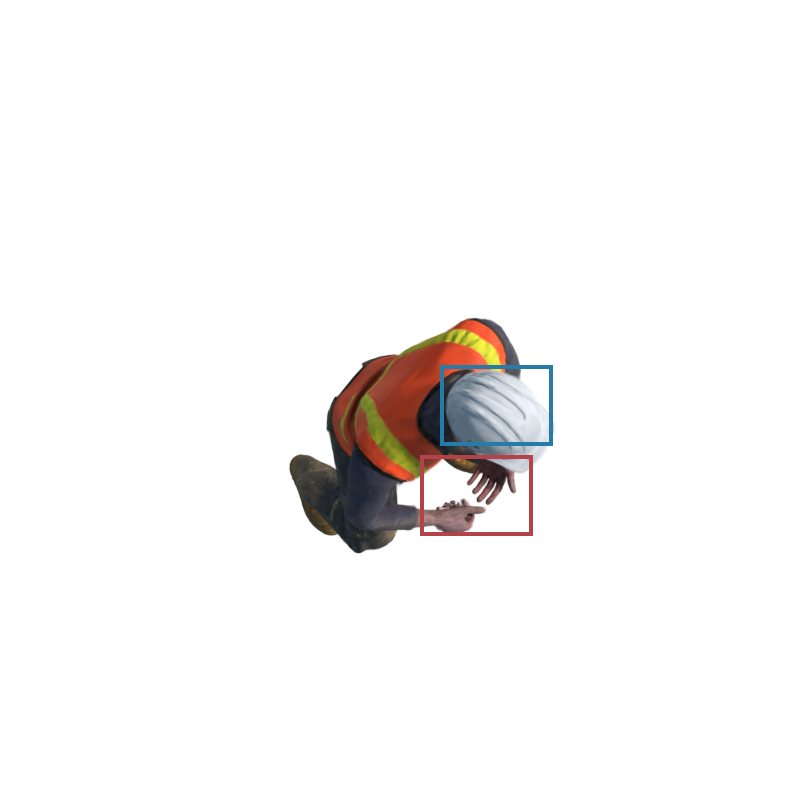} \\ &
    \includegraphics[width=0.160\linewidth]{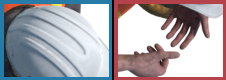} &
    \includegraphics[width=0.160\linewidth]{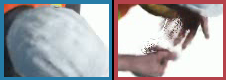} &
    \includegraphics[width=0.160\linewidth]{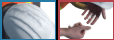} &
    \includegraphics[width=0.160\linewidth]{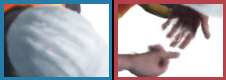} &
    \includegraphics[width=0.160\linewidth]{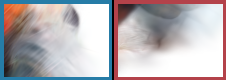} &
    \includegraphics[width=0.160\linewidth]{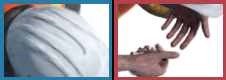} \\
    \raisebox{2em}{\rotatebox[origin=c]{90}{\footnotesize \textsc{Hook}}} & 
    \includegraphics[width=0.160\linewidth,trim={140px 160px 140px 120px},clip]{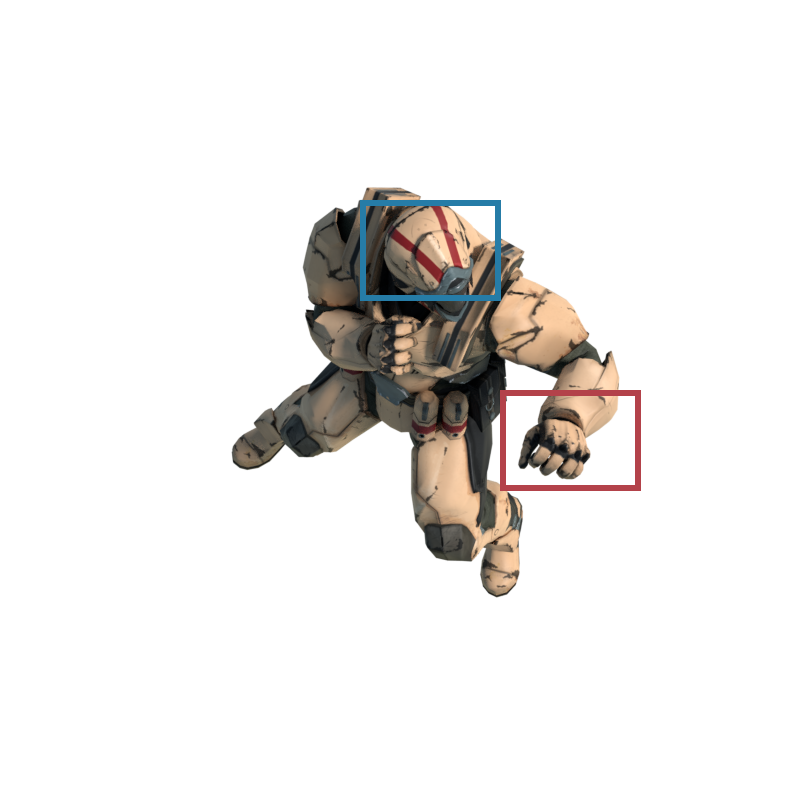} &
    \includegraphics[width=0.160\linewidth,trim={140px 160px 140px 120px},clip]{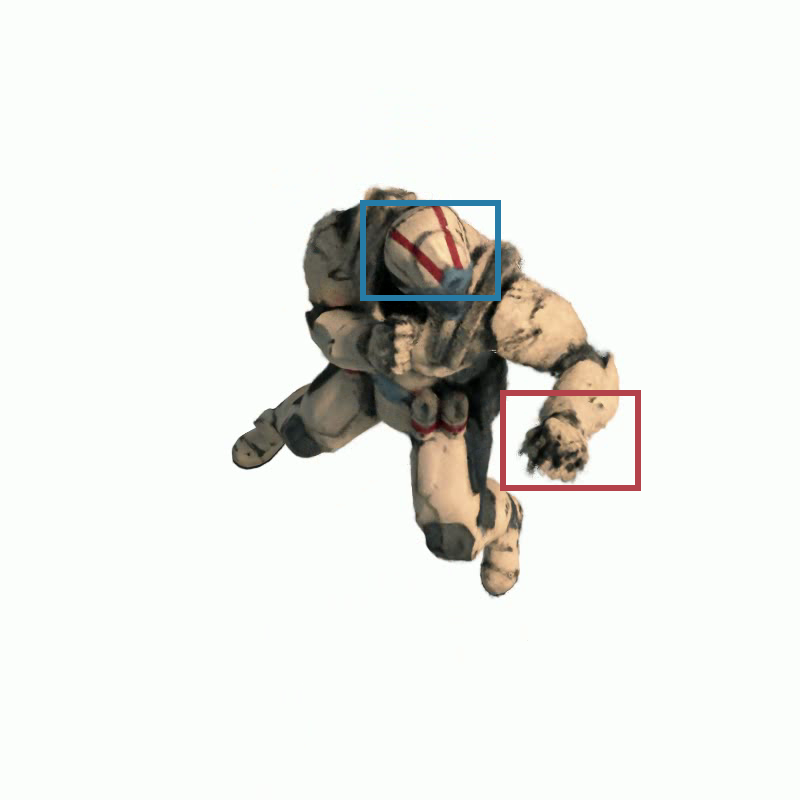} &
    \includegraphics[width=0.160\linewidth,trim={70px 80px 70px 60px},clip]{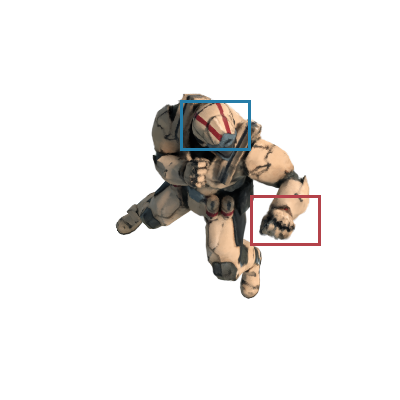} &
    \includegraphics[width=0.160\linewidth,trim={140px 160px 140px 120px},clip]{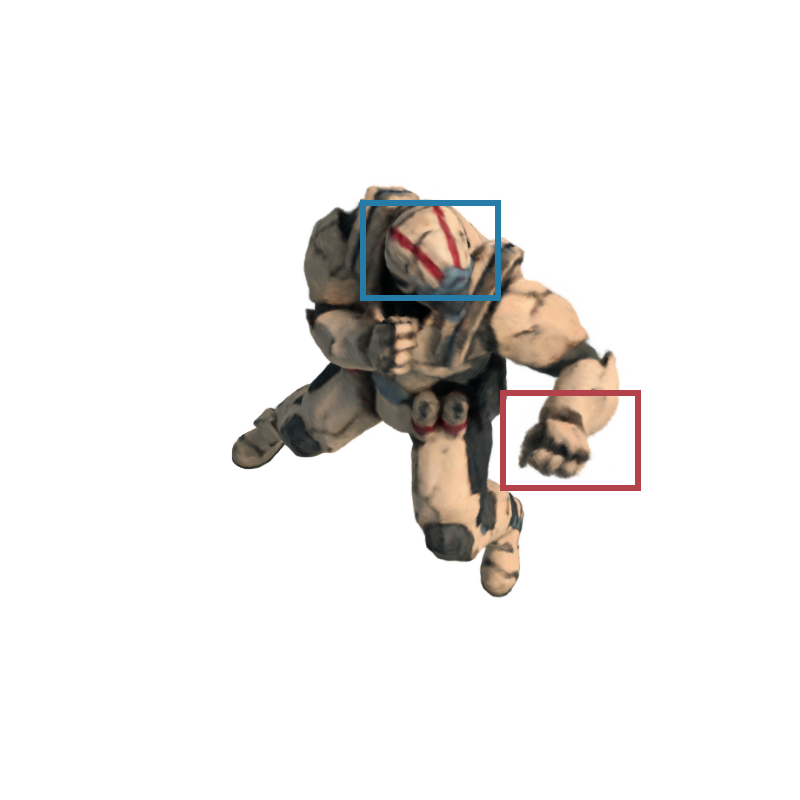} &
    \includegraphics[width=0.160\linewidth,trim={140px 160px 140px 120px},clip]{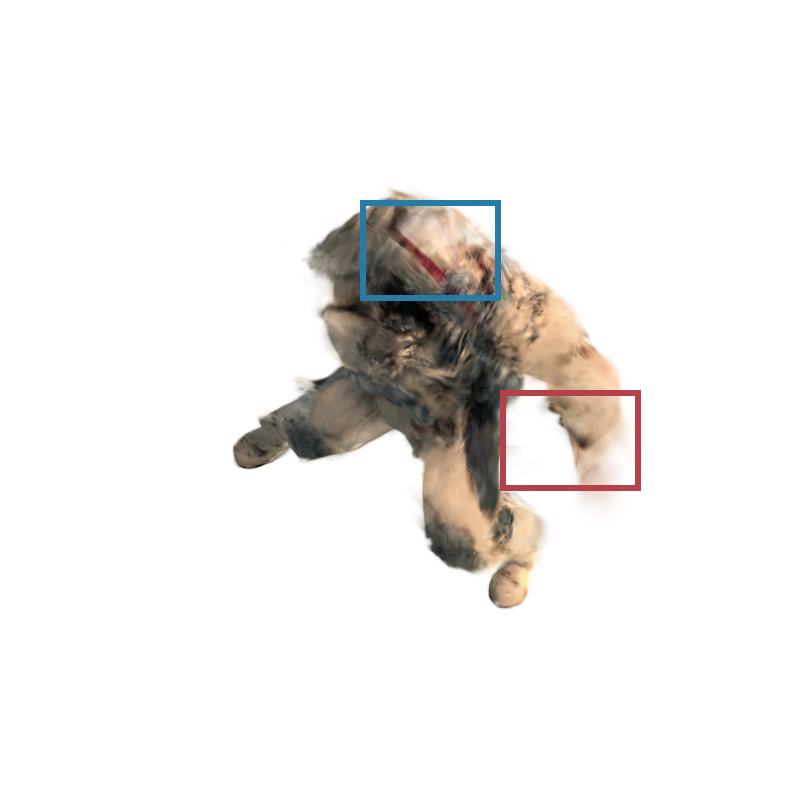} &
    \includegraphics[width=0.160\linewidth,trim={140px 160px 140px 120px},clip]{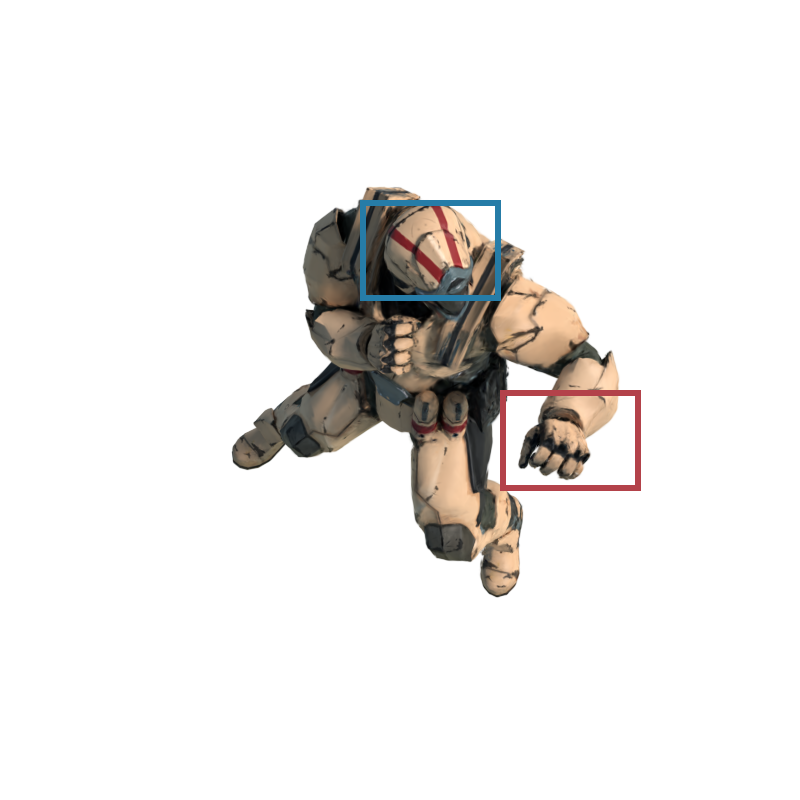} \\ &
    \includegraphics[width=0.160\linewidth]{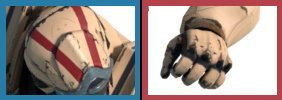} &
    \includegraphics[width=0.160\linewidth]{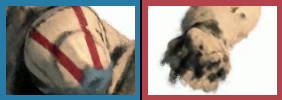} &
    \includegraphics[width=0.160\linewidth]{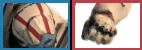} &
    \includegraphics[width=0.160\linewidth]{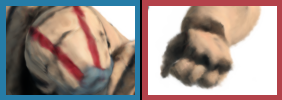} &
    \includegraphics[width=0.160\linewidth]{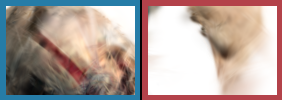} &
    \includegraphics[width=0.160\linewidth]{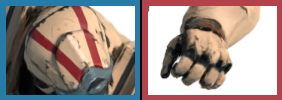} \\
    \raisebox{1em}{\smash{\rotatebox[origin=c]{90}{\scriptsize \textsc{Bouncing Balls}}}} & 
    \includegraphics[width=0.160\linewidth,trim={40px 160px 40px 100px},clip]{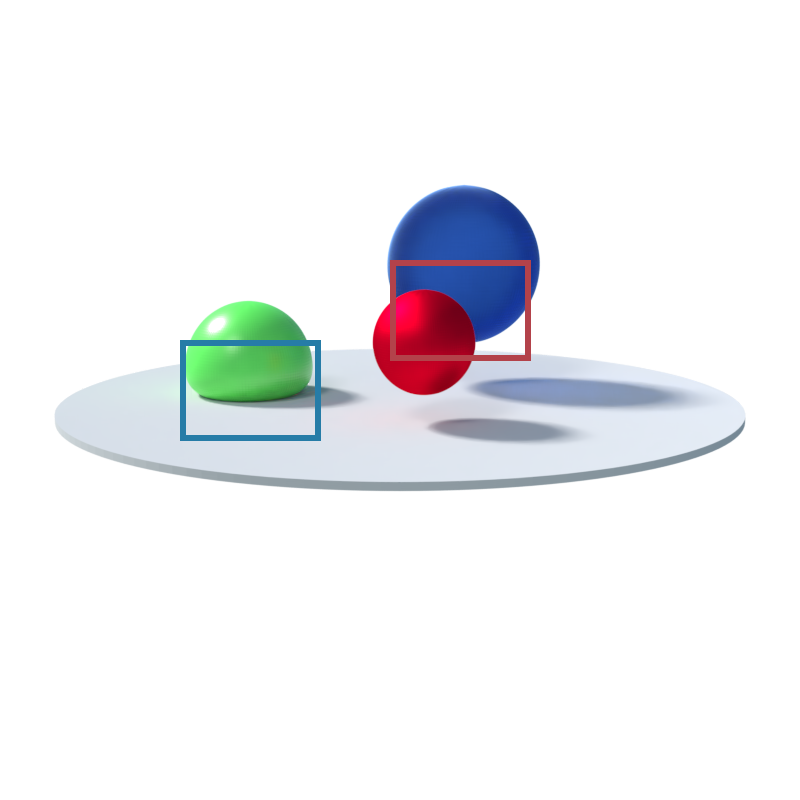} &
    \includegraphics[width=0.160\linewidth,trim={40px 160px 40px 100px},clip]{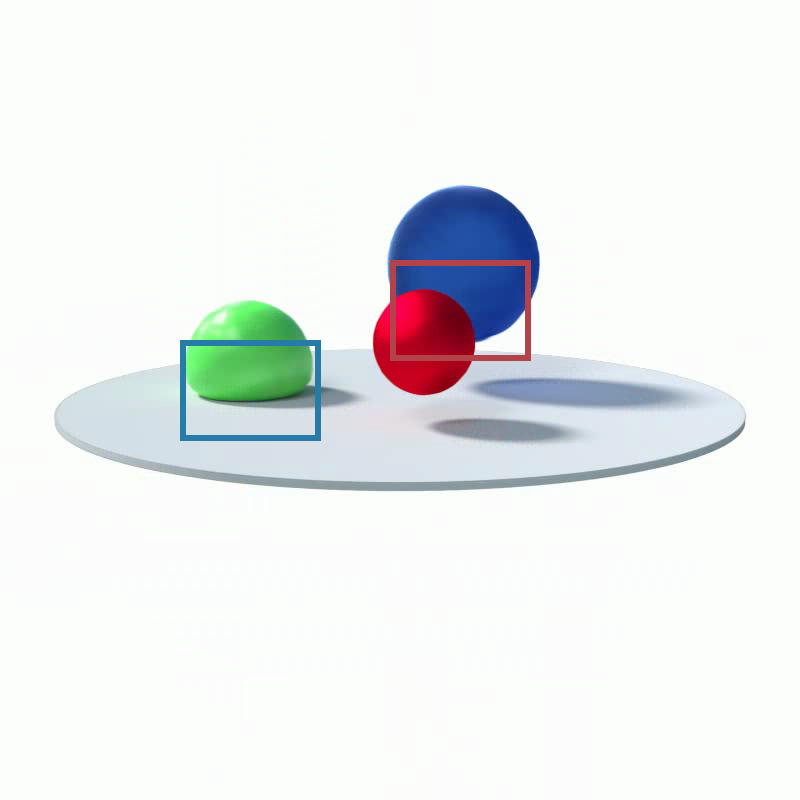} &
    \includegraphics[width=0.160\linewidth,trim={20px 80px 20px 50px},clip]{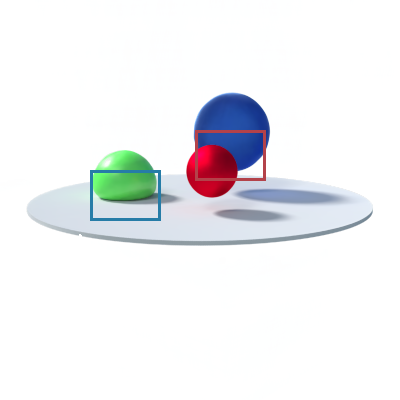} &
    \includegraphics[width=0.160\linewidth,trim={40px 160px 40px 100px},clip]{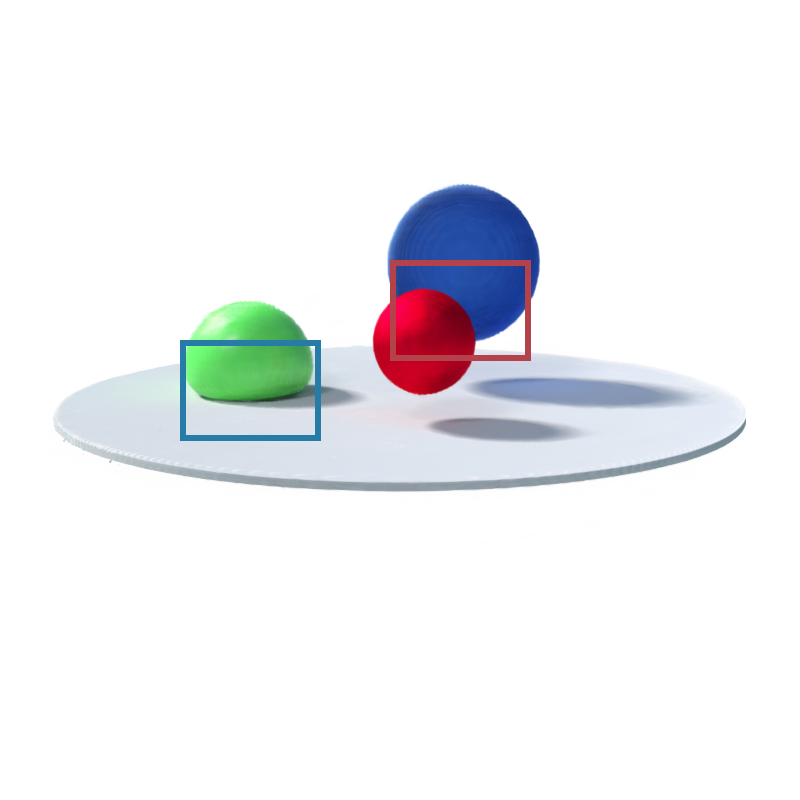} &
    \includegraphics[width=0.160\linewidth,trim={40px 160px 40px 100px},clip]{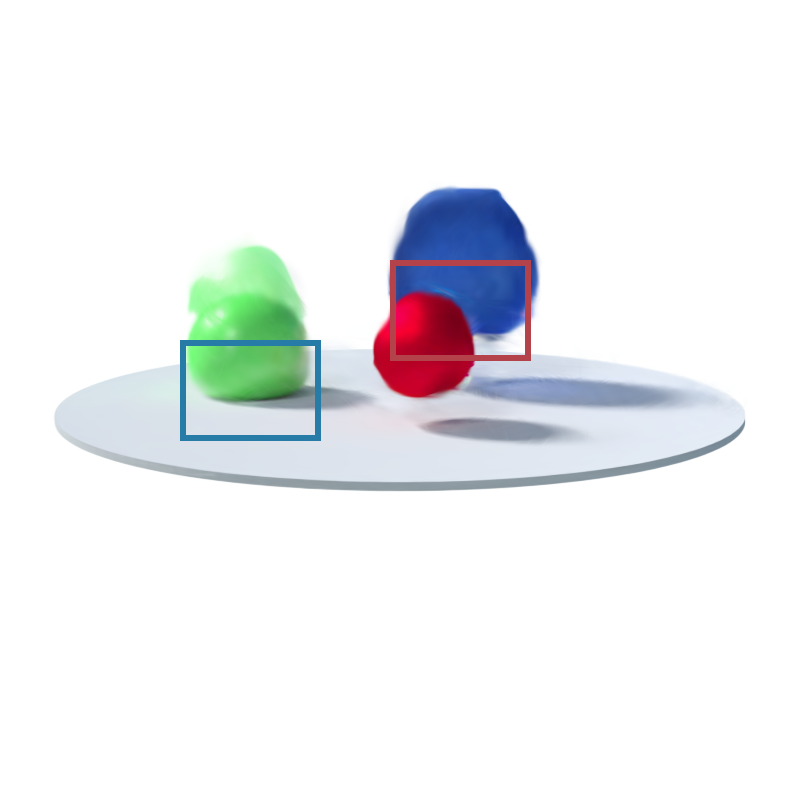} &
    \includegraphics[width=0.160\linewidth,trim={40px 160px 40px 100px},clip]{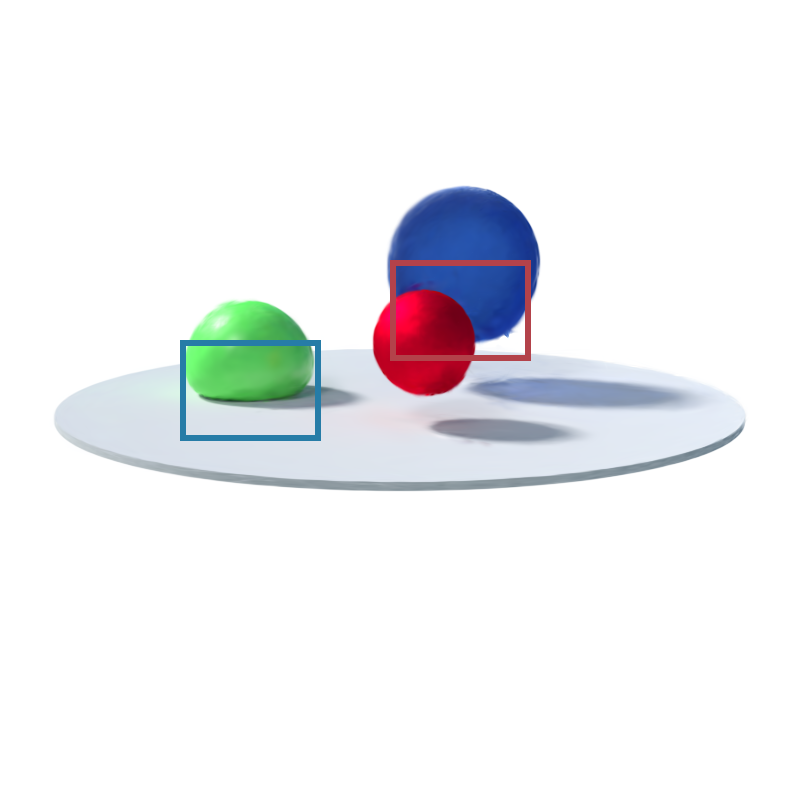} \\ &
    \includegraphics[width=0.160\linewidth]{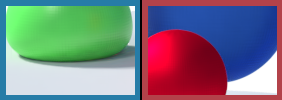} &
    \includegraphics[width=0.160\linewidth]{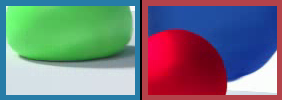} &
    \includegraphics[width=0.160\linewidth]{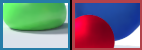} &
    \includegraphics[width=0.160\linewidth]{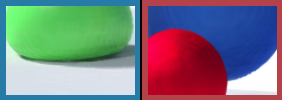} &
    \includegraphics[width=0.160\linewidth]{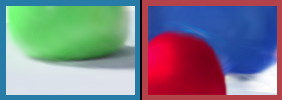} &
    \includegraphics[width=0.160\linewidth]{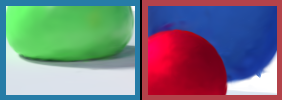} \\
    \raisebox{1.4em}{\smash{\rotatebox[origin=c]{90}{\footnotesize \textsc{Jumping Jacks}}}} &
    \includegraphics[width=0.160\linewidth,trim={50px 40px 50px 40px},clip]{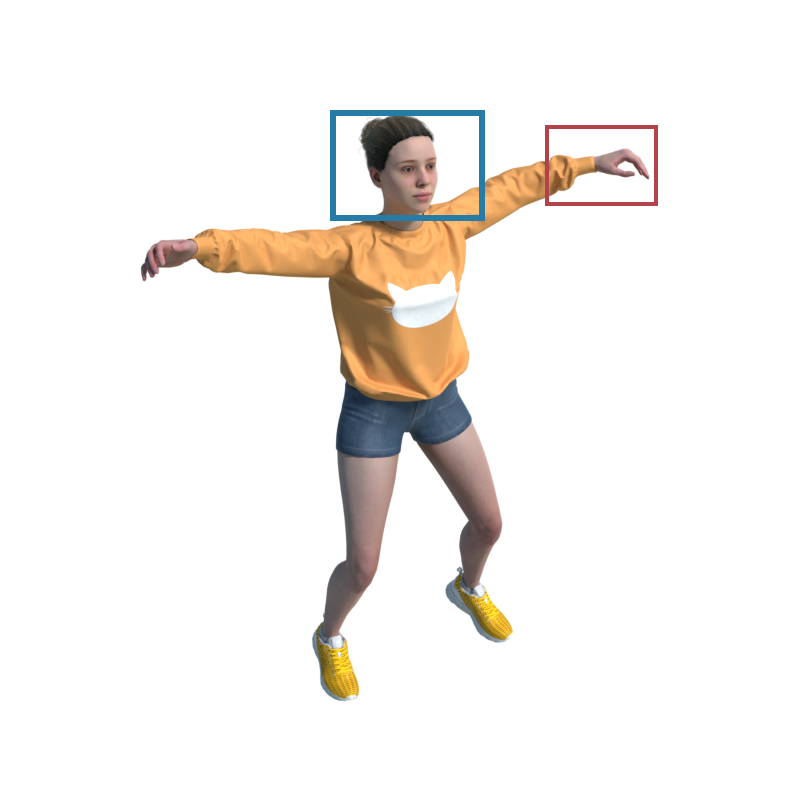} &
    \includegraphics[width=0.160\linewidth,trim={50px 40px 50px 40px},clip]{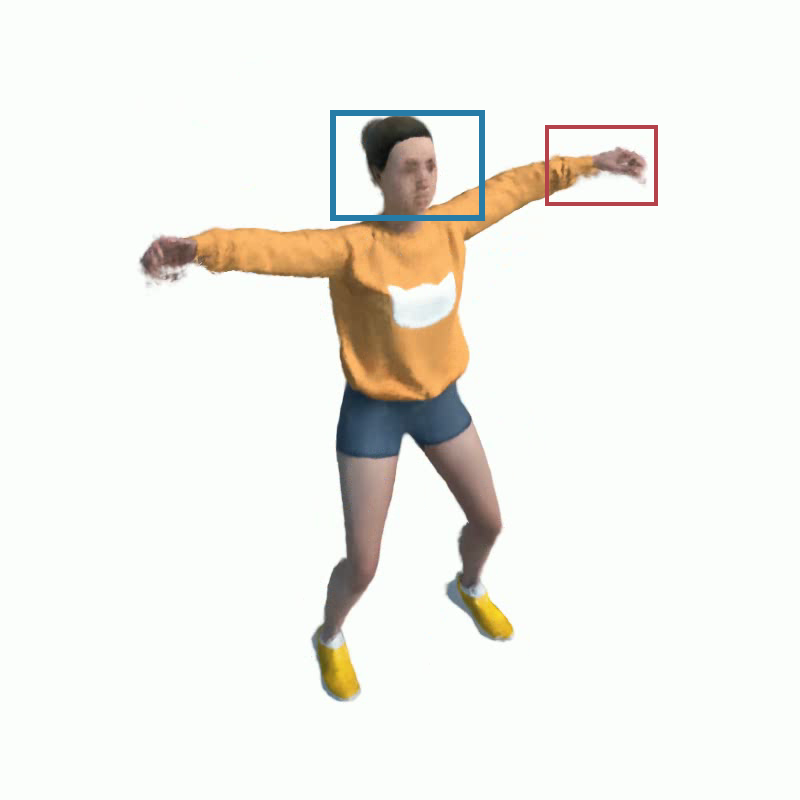} &
    \includegraphics[width=0.160\linewidth,trim={25px 20px 25px 20px},clip]{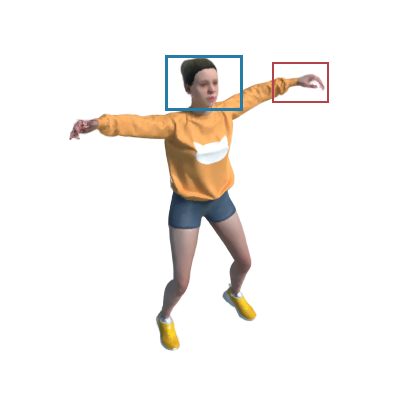} &
    \includegraphics[width=0.160\linewidth,trim={50px 40px 50px 40px},clip]{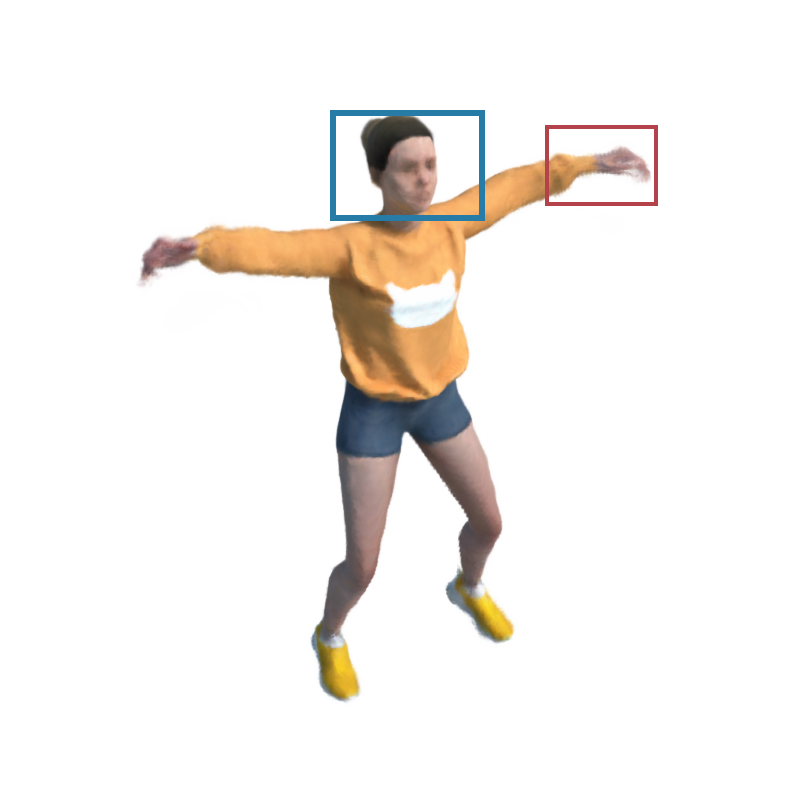} &
    \includegraphics[width=0.160\linewidth,trim={50px 40px 50px 40px},clip]{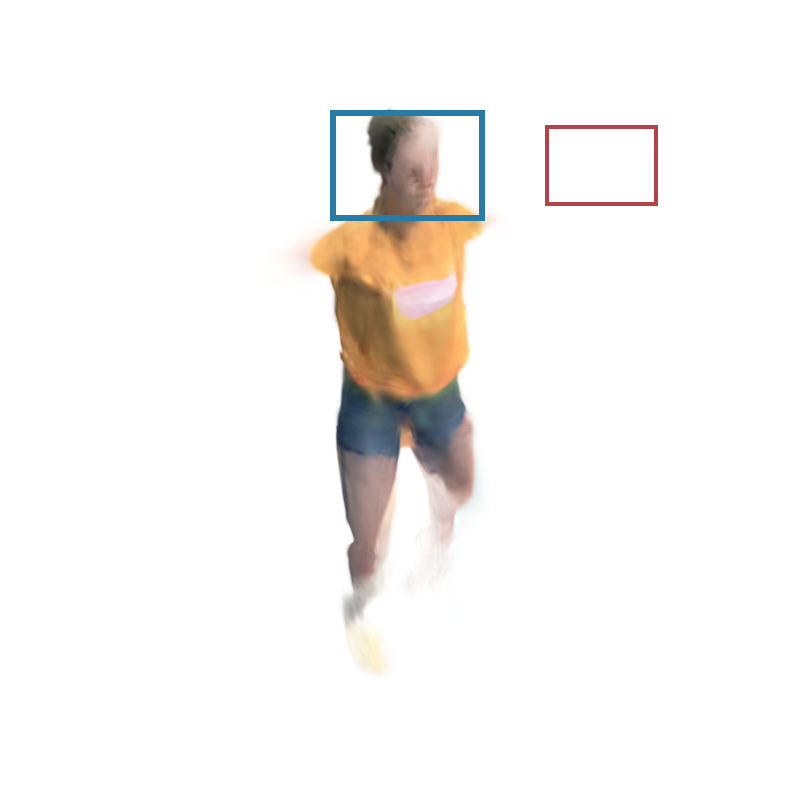} &
    \includegraphics[width=0.160\linewidth,trim={50px 40px 50px 40px},clip]{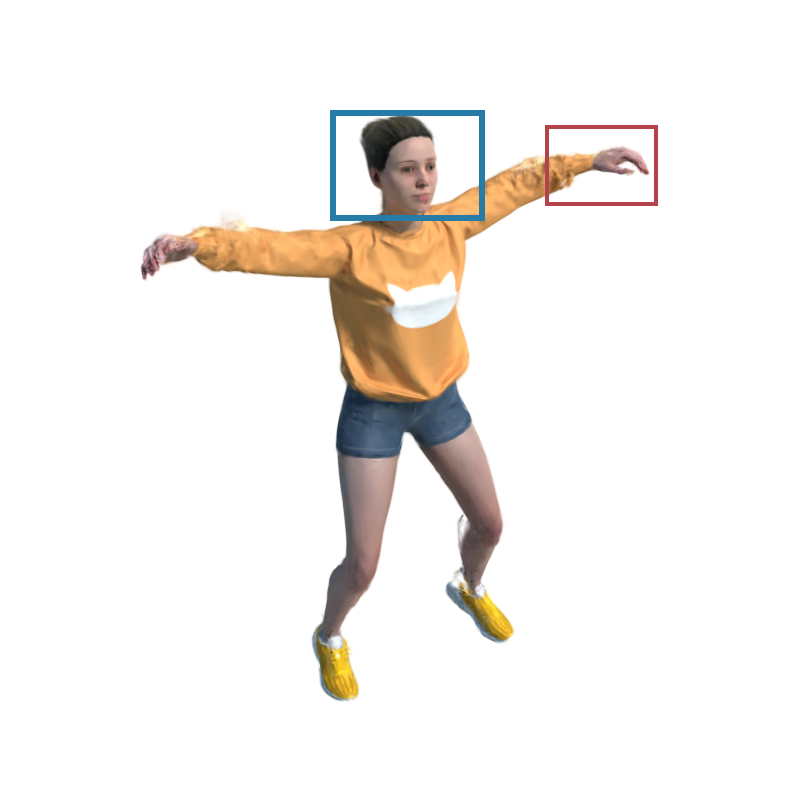} \\ &
    \includegraphics[width=0.160\linewidth]{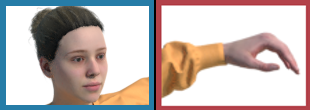} &
    \includegraphics[width=0.160\linewidth]{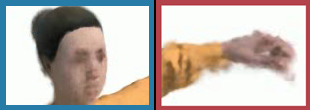} &
    \includegraphics[width=0.160\linewidth]{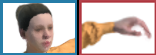} &
    \includegraphics[width=0.160\linewidth]{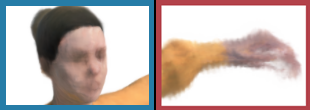} &
    \includegraphics[width=0.160\linewidth]{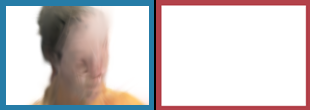} &
    \includegraphics[width=0.160\linewidth]{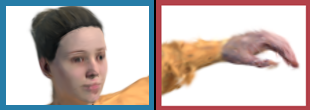} 
  \end{tabular}\egroup
\caption{Qualitative comparison on D-NeRF~\cite{pumarola2021d}. We highlight the difference by zoom view. Our method achieves competitive visual quality with strong baselines. While our method successfully reconstructs intricate details like hands, it causes a blurred sphere shape.}\label{fig:dnerf}
\end{figure*}

\section{Experiment}

\subsection{Evaluation data}
We evaluate our compact dynamic Gaussian representation using three dynamic scene datasets: a synthetic one D-NeRF~\cite{pumarola2021d}, and two real ones DyNeRF~\cite{li2022neural} and HyperNeRF~\cite{park2021hypernerf}.

\noindent \textbf{D-NeRF dataset~\cite{pumarola2021d}.}
This dataset comprises eight videos of varying lengths, ranging from 50 to 200 frames per video. The camera setup is designed to mimic a monocular camera setting by teleporting between adjacent timesteps. The test views are from novel camera positions. 
We train and render at the resolution of $800 \times 800$.

\noindent \textbf{DyNeRF dataset~\cite{li2022neural}.}
The multi-camera dataset includes six 10-second videos captured at 30 FPS using 15--20 synchronized fixed cameras. For evaluation, a central camera is used, while training utilizes frames from the other cameras. The training and rendering resolution is set at 1,352$\times$1,014.

\noindent \textbf{HyperNeRF dataset~\cite{park2021hypernerf}.}
This dataset encompasses videos ranging from 8 to 15 seconds, captured at 15 FPS using two Pixel 3 phones. The training and rendering processes are conducted at a resolution of $540 \times 960$.

\begin{table}[tb]
\centering
\caption{Quantitative results on the D-NeRF dataset~\cite{pumarola2021d}. Our method performs competitively against NeRF approaches in terms of visual quality and achieves the fastest rendering speed among the methods with high quality. Results excepting FPS of ~\cite{TiNeuVox,fridovich2023k,gan2023v4d} are adopted from the original papers. The \best{best} and \sbest{second best} scores among competing methods are highlighted.}\label{tb:dnerf}
\resizebox{0.7\linewidth}{!}{
\begin{tabular}{lcccccc}\toprule
& PSNR$\uparrow$ & MS-SSIM$\uparrow$  & LPIPS$\downarrow$ & FPS $\uparrow$ & Train Time $\downarrow$ & Mem$\downarrow$ \\\midrule
TiNeuVox-S~\cite{TiNeuVox} &  30.75 & 0.96 & 0.07  & 0.32 & \best{8 mins} & \best{8MB}\\
TiNeuVox-B~\cite{TiNeuVox} & \sbest{32.67} & \sbest{0.97} & \sbest{0.04}  & 0.13 & \sbest{28 mins} & \sbest{48MB} \\
K-Planes~\cite{fridovich2023k} & 31.61 & \sbest{0.97} & - & 0.54 & 52 mins & $\sim$497MB \\ 
V4D~\cite{gan2023v4d} & \best{33.72} & \best{0.98} & \best{0.02} & \sbest{1.47} & 6.9 hrs & 1.2GB \\
3DGS~\cite{kerbl20233d} & 20.51 & 0.89 & 0.07 & 170 & 6 mins & $\sim$50MB \\
D-3DGS & 17.22 & 0.81 & 0.13 & 173 & 15 mins & $\sim$913MB \\
Ours & 32.19 &  \sbest{0.97} & \sbest{0.04} & \best{150} & \best{8 mins} & $\sim$159MB \\
\bottomrule
\end{tabular}}
\end{table}

\begin{table}[tb]
\centering
\caption{Quantitative results on the DyNeRF datasets~\cite{li2022neural}. 
Results excepting FPS of ~\cite{fridovich2023k,gan2023v4d} are adopted from the original papers. The \best{best} and \sbest{second best} scores among competing methods (excepting 3DGS) are highlighted. While our method is match for NeRF in rendering quality, our method is match for 3DGS in rendering speed. Besides, our method is 20 times more compact than Dynamic3DGaussians.
}\label{tb:dynerf}
\resizebox{0.74\linewidth}{!}{
\begin{tabular}{lcccccc}\toprule
& PSNR$\uparrow$ & MS-SSIM$\uparrow$  & LPIPS$\downarrow$ & FPS$\uparrow$ & Train\,Time$\downarrow$ & Mem$\downarrow$ \\\midrule
K-Planes~\cite{fridovich2023k} & \best{31.63} & \best{0.964} & - & 0.31 & 1.8 hrs & \best{$\sim$309MB} \\ 
V4D &  28.96 & 0.937 & \sbest{0.17} & 0.11 & 4 hrs & 1.2GB  \\
3DGS~\cite{kerbl20233d} & 20.94 & 0.800 & 0.29 & 109 & 20 mins & $\sim$198MB \\
D-3DGS & 24.36 & 0.834 & 0.25 & \best{119} & \best{51 mins}  & $\sim$2.3GB \\
Dynamic3DGaussians~\cite{luiten2023dynamic} & 27.79 & 0.869 & 0.23 & 51 & 2.1 hrs & $\sim$6.6GB\\
Ours & \sbest{30.46} &  \sbest{0.955} & \best{0.15} & \sbest{118} & \sbest{1 hrs} & \sbest{$\sim$338MB} \\
\bottomrule
\end{tabular}}
\end{table}

\subsection{Implementation details}

We adhere to the experimental setup in the 3DGS paper~\cite{kerbl20233d}. The number of approximation terms of Gaussian centers $L$ is set to 2 for the D-NeRF dataset. For the DyNeRF and HyperNeRF datasets, $L$ is set to be 5 from preliminary experiments. 
Our two-stage optimization process begins with an initial fitting of parameters, excluding the coefficients for Gaussian position and variance. This initial stage spans 3,000 iterations and utilizes all training views in a static setting. Subsequently, we engage in a dynamic stage, adjusting all Gaussian parameters in 27,000 iterations. The entire optimization process encompasses 30,000 iterations.
Following~\cite{kerbl20233d}, $\lambda$ is set to be 0.2.
We set the flow loss weight $\lambda_{\rm flow}$ at 1,000 and acquire ground truth flow through RAFT pretrained on the Sintel dataset~\cite{Butler2012}. All experiments are conducted on a single RTX A6000 GPU.

\begin{table*}[t]
\caption{Quantitative results on the HyperNeRF dataset~\cite{park2021hypernerf}. Our method demonstrates competitive performance in rendering quality across all scenes, surpassing the compared methods in rendering speed. Furthermore, our method is not inferior to the compared methods in training time and memory size.}\label{tb:hypernerf}
\resizebox{\textwidth}{!}{
\begin{tabular}{lccccccccccccc}\toprule
& \multirow{2}{*}{\!FPS$\uparrow$\!} &\multirow{2}{*}{\centering Train\,Time$\downarrow$} & \!\multirow{2}{*}{\centering Mem$\downarrow$}\! & \multicolumn{2}{c}{\textsc{Broom}} &  \multicolumn{2}{c}{\textsc{3D\,Printer}} & \multicolumn{2}{c}{\textsc{Chicken}} & \multicolumn{2}{c}{\!\textsc{Peel\,Banana}\!} & \multicolumn{2}{c}{\textbf{Mean}}\\\cmidrule(l{2pt}r{2pt}){5-6}\cmidrule(l{2pt}r{2pt}){7-8}\cmidrule(l{2pt}r{2pt}){9-10} \cmidrule(l{2pt}r{2pt}){11-12} \cmidrule(l{2pt}r{2pt}){13-14}
& & & &\footnotesize \!PSNR$\uparrow$\! &\footnotesize \!SSIM$\uparrow$\!  &\footnotesize \!PSNR$\uparrow$\! &\footnotesize \!SSIM$\uparrow$\! &\footnotesize \!PSNR$\uparrow$\! &\footnotesize \!SSIM$\uparrow$\! &\footnotesize \!PSNR$\uparrow$\! &\footnotesize \!SSIM$\uparrow$\! &\footnotesize \!PSNR$\uparrow$\! &\footnotesize \!SSIM$\uparrow$\! \\\midrule
HyperNeRF~\cite{park2021hypernerf} & \sbest{0.36} & \!\!48 hrs\dag\!\! & \!\!\sbest{15MB}\!\! & 19.3 & 0.591 & 20.0 & 0.821 & 26.9 & \best{0.948} & 23.3 & \sbest{0.896} & 22.2 & 0.811 \\
TiNeuVox-B~\cite{TiNeuVox} & 0.14 & \!\!\best{30 mins}\!\! & \!\!\sbest{48MB}\!\! & \sbest{21.5} & \sbest{0.686} & 22.8  & \sbest{0.841} & \sbest{28.3} & \sbest{0.947} & 24.4 & 0.873 & 24.3 & \sbest{0.837} \\ 
V4D~\cite{gan2023v4d} & 0.15 & \!\!7 hrs\!\! & 1.2GB & \best{22.1} & 0.669 & \sbest{23.2} & 0.835 & \best{28.4} & 0.929 & \sbest{25.2} & 0.873 & \sbest{24.7} & 0.827 \\ 
Ours & \best{188} & \sbest{1 hrs} & \!\!$\sim$720MB\!\! & \best{22.1} & \best{0.789} & \best{25.5} & \best{0.919} & \sbest{28.3} & {0.934} & \best{26.6} & \best{0.920} & \best{25.6} & \best{0.890} \\\bottomrule
\end{tabular}}
{\begin{spacing}{0.9}\scriptsize \dag Train time of HyperNeRF~\cite{park2021hypernerf} is estimated from their paper's descriptions. Originally reported as 8 hours on 4 TPU v4s~\cite{jouppi2023tpu}, the TPU v4 is slightly faster than the A100 GPU, and the A100 GPU is at least 1.5 times faster than the A6000 GPU. \end{spacing}}
\end{table*}

\begin{figure*}[t]
  \centering
  \bgroup 
  \def\arraystretch{0.1} 
  \setlength\tabcolsep{0.2pt}
  \begin{tabular}{ccc}
    Ground Truth & K-Planes~\cite{fridovich2023k} & Ours \\\\
    \includegraphics[width=0.32\linewidth,trim={0px 0px 0px 200px},clip]{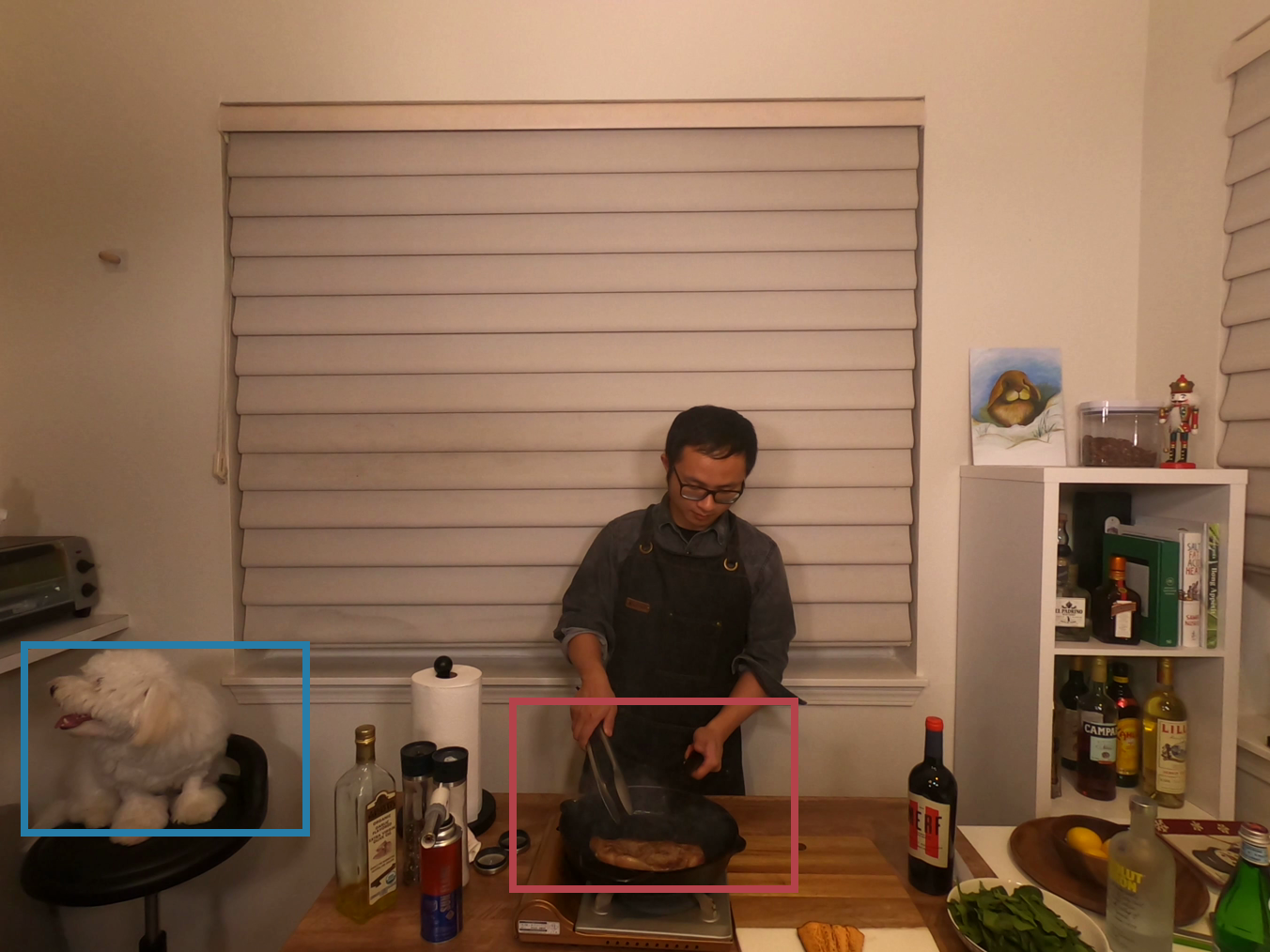} &
    \includegraphics[width=0.32\linewidth,trim={0px 0px 0px 200px},clip]{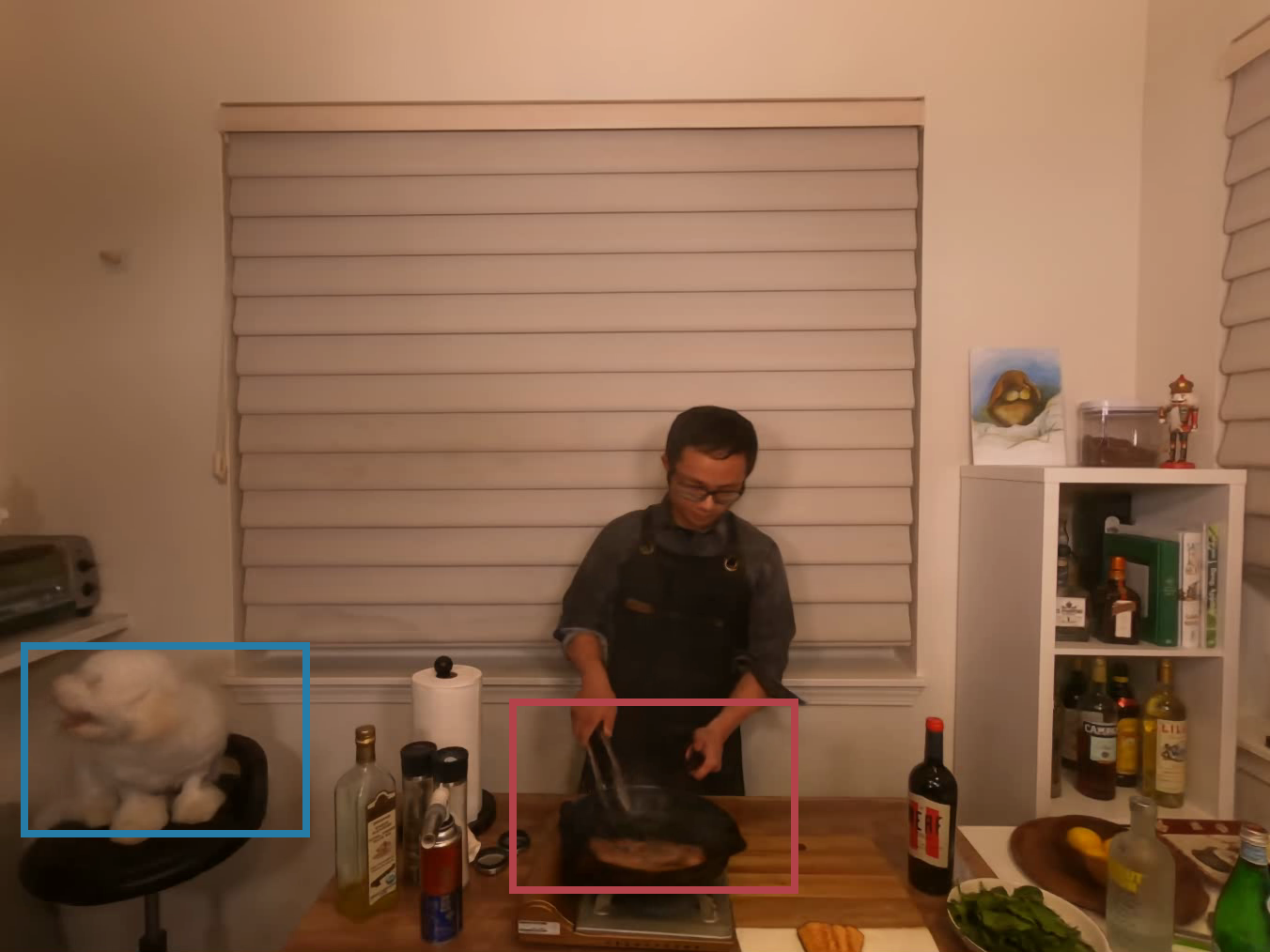} &
    \includegraphics[width=0.32\linewidth,trim={0px 0px 0px 200px},clip]{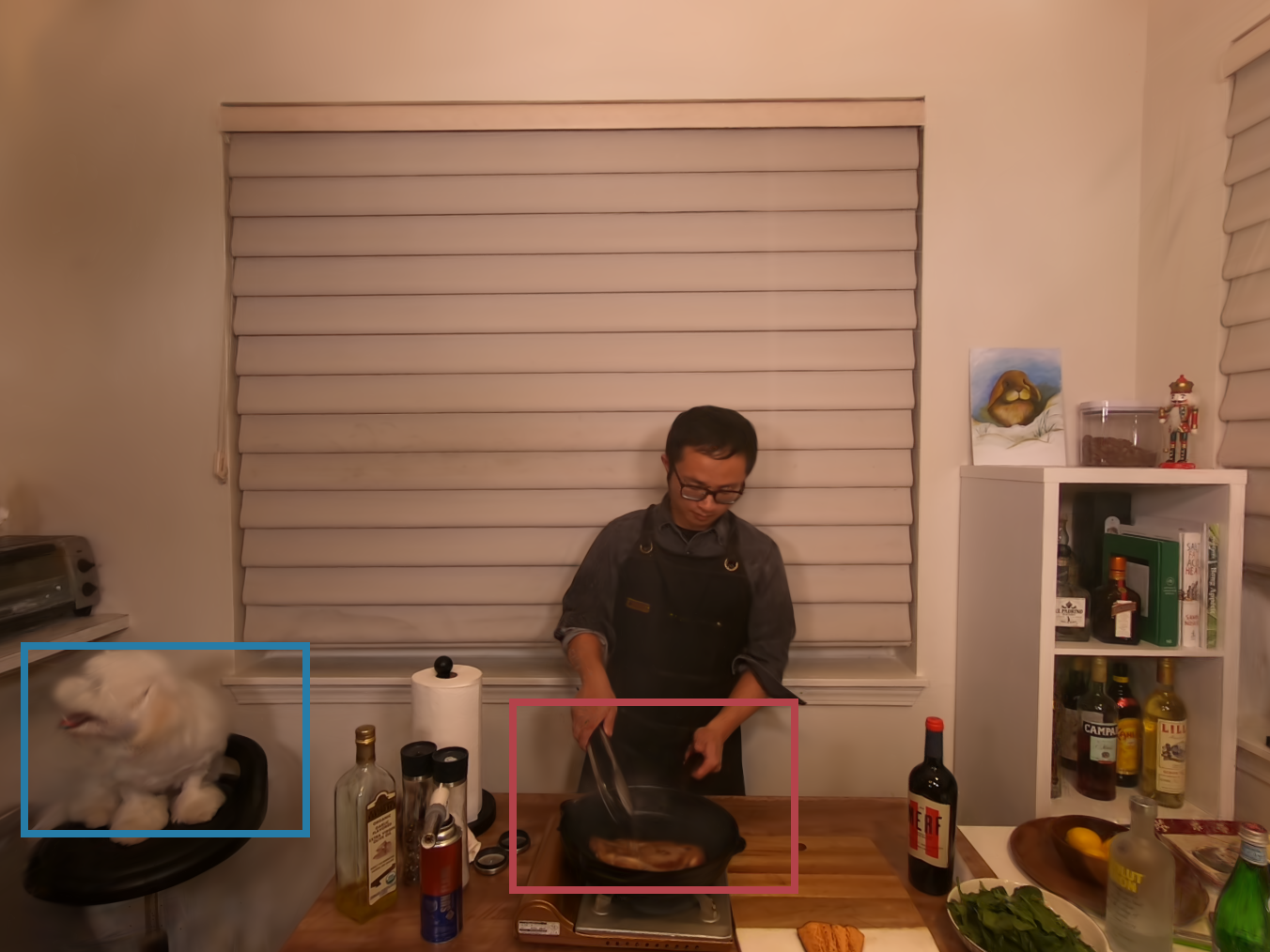} \\
    \includegraphics[width=0.32\linewidth,trim={0px 0px 0px 00px},clip]{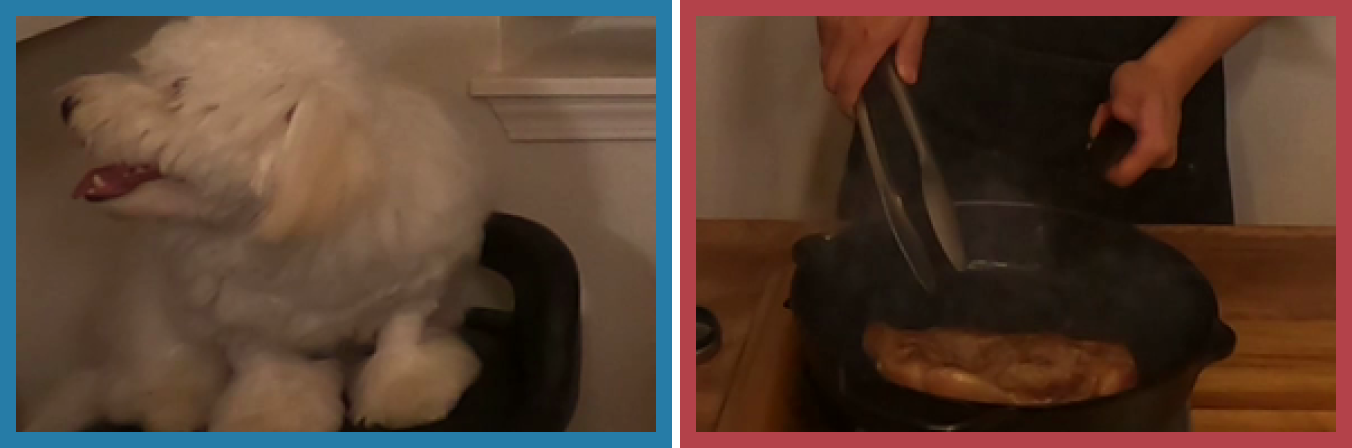} &
    \includegraphics[width=0.32\linewidth,trim={0px 0px 0px 00px},clip]{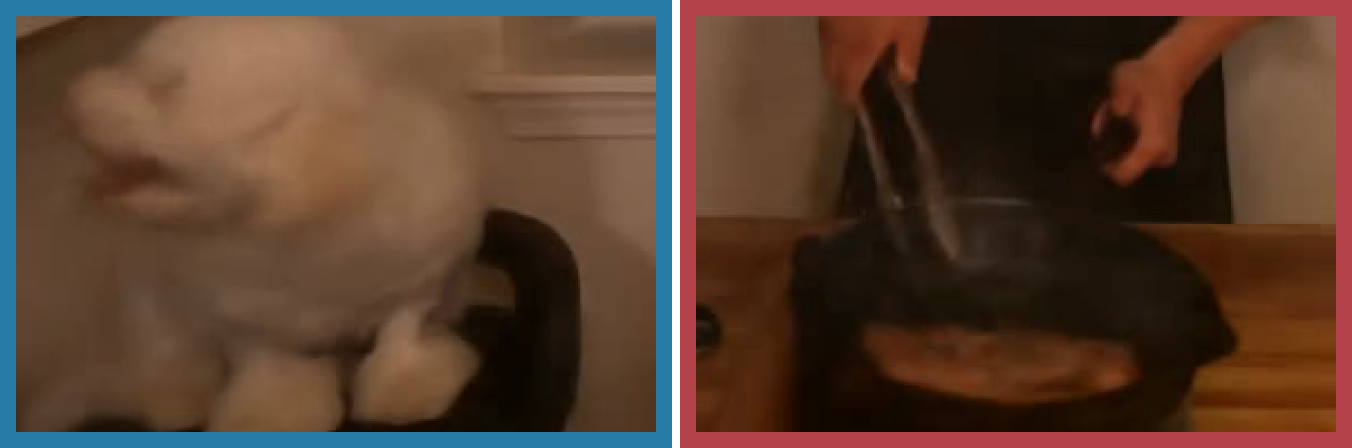} &
    \includegraphics[width=0.32\linewidth,trim={0px 0px 0px 00px},clip]{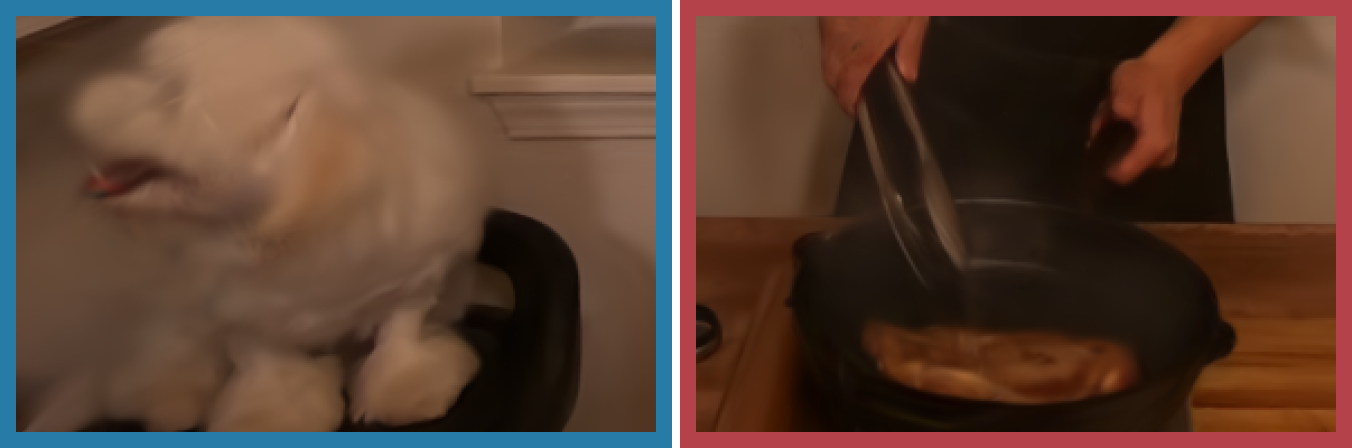} \\
  \end{tabular}\egroup
\caption{Qualitative comparison on the DyNeRF dataset~\cite{li2022neural}. 
The differences are zoomed in.
}\label{fig:dynerf}
\end{figure*}

\begin{figure}[t]
  \centering
  \bgroup 
  \def\arraystretch{0.1} 
  \setlength\tabcolsep{0.2pt}
  \begin{tabular}{ccccccc}
    Ground Truth & V4D~\cite{gan2023v4d} & Ours && Ground Truth & V4D~\cite{gan2023v4d} & Ours \\\\
    \includegraphics[width=0.16\linewidth]{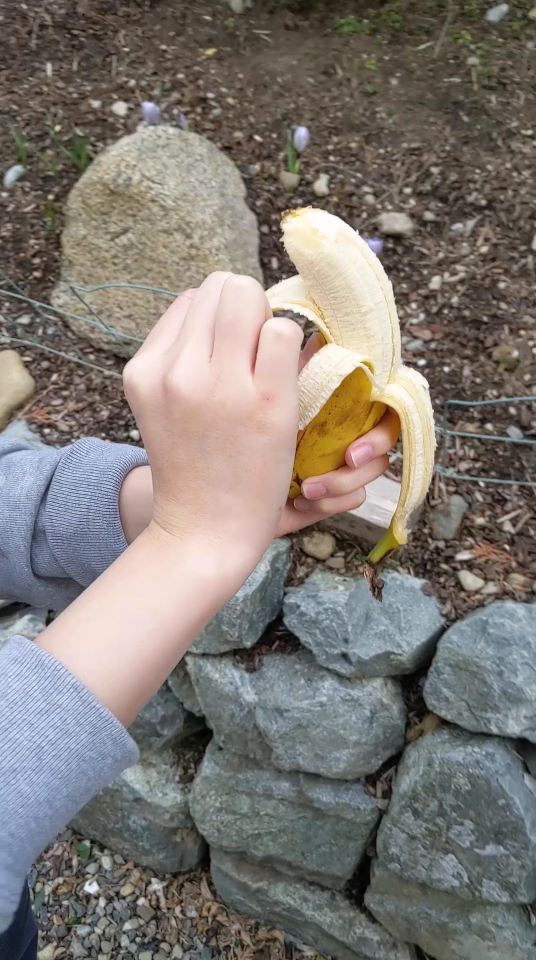} &
    \includegraphics[width=0.16\linewidth]{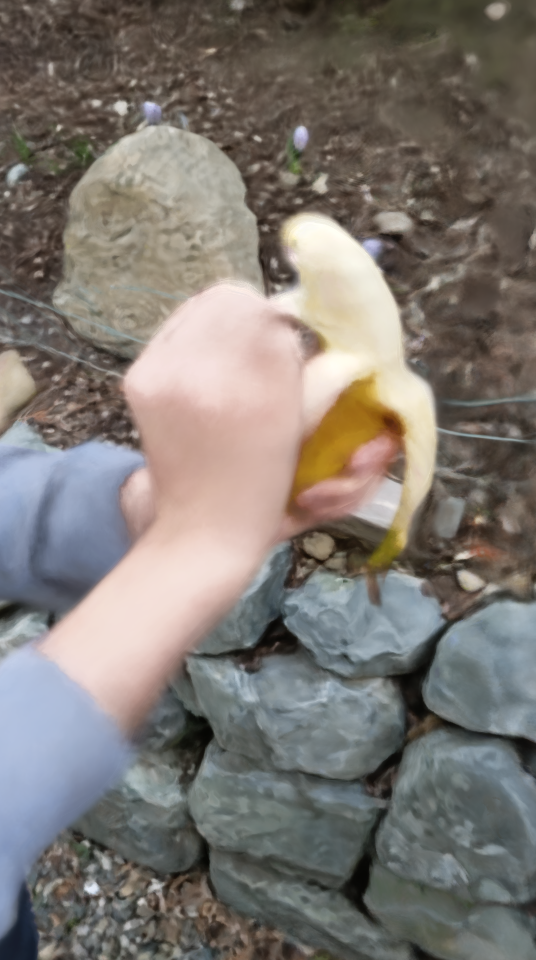} &
    \includegraphics[width=0.16\linewidth]{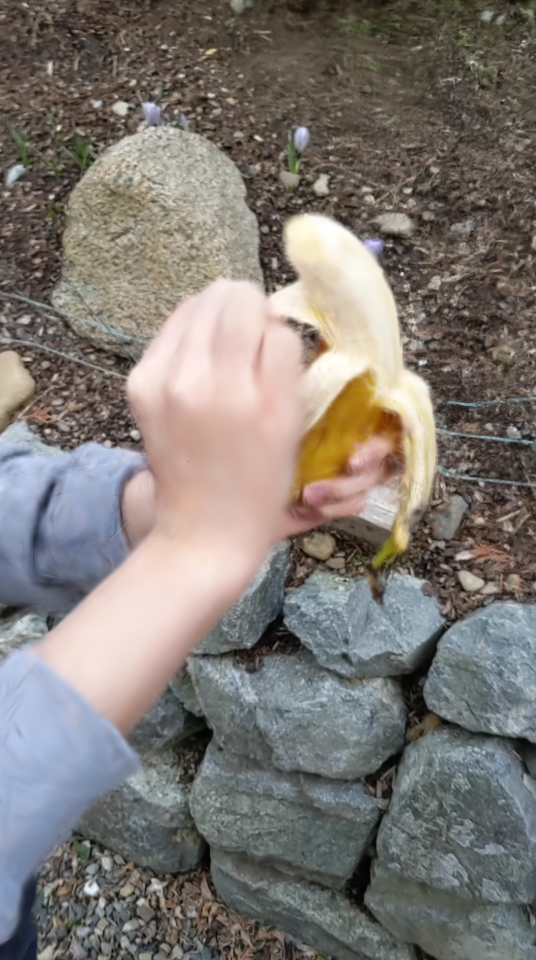} &&
    \includegraphics[width=0.16\linewidth]{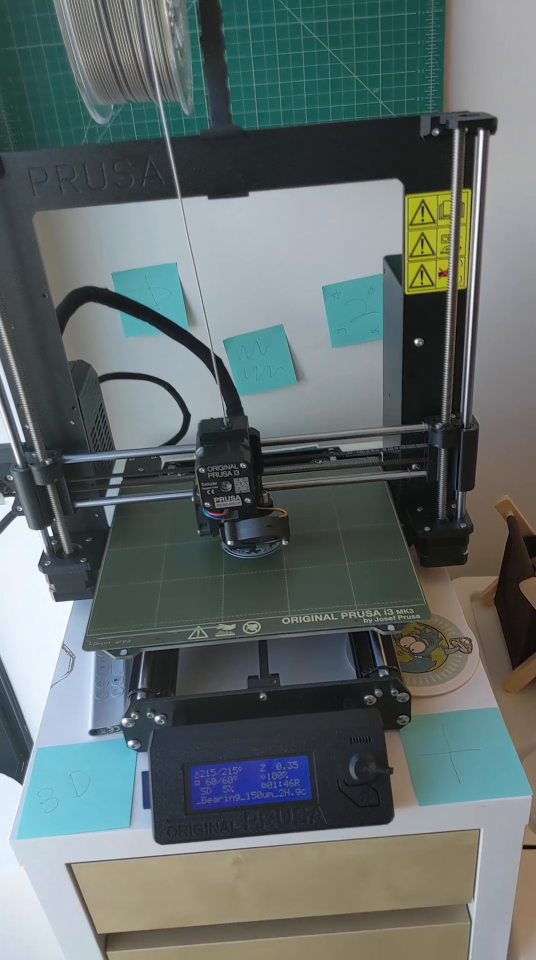} &
    \includegraphics[width=0.16\linewidth]{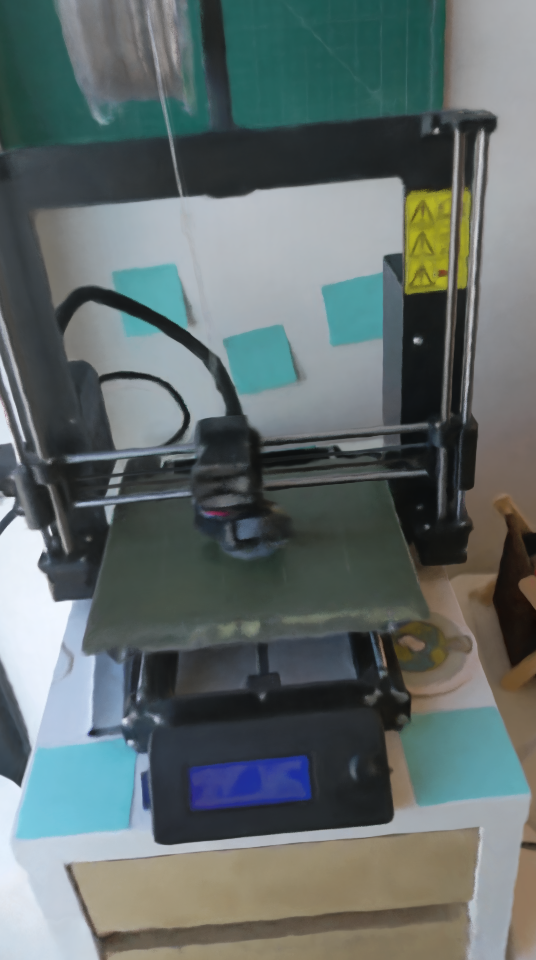} &
    \includegraphics[width=0.16\linewidth]{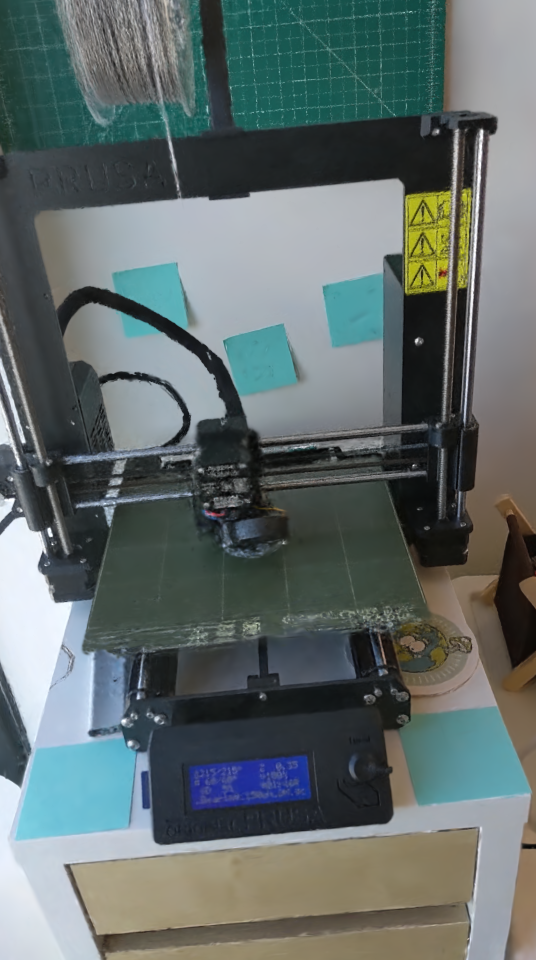} \\
  \end{tabular}\egroup
\caption{Qualitative comparison on HyperNeRF~\cite{park2021hypernerf}. Our method offers sharp results.}\label{fig:hypernerf} %
\end{figure}

\subsection{Evaluation setup}
\noindent \textbf{Compared methods.}
We  benchmark our method against several SoTA methods, including TiNeuVox~\cite{TiNeuVox}, K-Planes~\cite{fridovich2023k}, V4D~\cite{gan2023v4d}, HyperNeRF~\cite{park2021hypernerf}, 3D Gaussian Splatting (3DGS)~\cite{kerbl20233d}, Dynamic3DGaussians~\cite{luiten2023dynamic}, and a D-3DGS baseline. D-3DGS is the dynamic extension of 3DGS, which stores both position and rotation for each timestep.

\noindent \textbf{Evaluation metrics.}
We assess the methods using various metrics, including PSNR~\cite{huynh2008scope}, SSIM~\cite{wang2004image}, LPIPS~\cite{zhang2018unreasonable}, FPS, Training time, and memory used to store optimized parameters. Memory consumption includes 3D Gaussian parameters, voxel/plane representation, neural network parameters, and so on.

\begin{table}[tb]
\centering
\small
\caption{Per-scene quantitative comparison on D-NeRF scenes of different $L$, which stands for the number of harmonics term in the Fourier approximation, and other design choices. The highest mean score is achieved with $L=2$, but increasing the complexity L (the number of coefficients) improves visual quality in some scenes (\textsc{Jumping Jacks} and \textsc{T-Rex}). The spline approximations bring marginal advantages in some scenes, but slower rendering.
The time-varying scale (the last row) also gives minor gains in some cases and increases the memory size. The setting reported in \cref{fig:dnerf} is highlighted in a gray background.} \label{tb:ablation_l}
\resizebox{\linewidth}{!}{
\scriptsize
\begin{tabular}{lcccccccccccccccccc}\toprule
& \multicolumn{2}{c}{\textsc{Stand\,Up}} &\multicolumn{2}{c}{\textsc{Jacks}} & \multicolumn{2}{c}{\textsc{Balls}} & \multicolumn{2}{c}{\textsc{Lego}} & \multicolumn{2}{c}{\textsc{Warrior}} & \multicolumn{2}{c}{\textsc{Hook}} & \multicolumn{2}{c}{\textsc{T-Rex}} & \multicolumn{2}{c}{\textsc{Mutant}} & \multicolumn{2}{c}{\textbf{Mean}}\\\cmidrule(l{2pt}r{2pt}){2-3}\cmidrule(l{2pt}r{2pt}){4-5}\cmidrule(l{2pt}r{2pt}){6-7}\cmidrule(l{2pt}r{2pt}){8-9} \cmidrule(l{2pt}r{2pt}){10-11} \cmidrule(l{2pt}r{2pt}){12-13}\cmidrule(l{2pt}r{2pt}){14-15}\cmidrule(l{2pt}r{2pt}){16-17}\cmidrule(l{2pt}r{2pt}){18-19}
& \!PSNR\! & \!SSIM\!  & \!PSNR\! & \!SSIM\! & \!PSNR\! & \!SSIM\! & \!PSNR\! & \!SSIM\! & \!PSNR\! & \!SSIM\! & \!PSNR\! & \!SSIM\! & \!PSNR\! & \!SSIM\! & \!PSNR\! & \!SSIM\! & \!PSNR\! & \!SSIM\! \\\midrule
$L=1$ & \best{40.21} & \best{0.994} & 27.22 & 0.952 & 30.27 & 0.972 & \best{24.26} & \best{0.940} & 32.42 & 0.937 & 32.84 & \sbest{0.980} & 25.15 & {0.957} & \best{38.04} & \best{0.994} & 31.30 & 0.965\\
\rowcolor{gray!20} $L=2$ & \sbest{39.10} & \sbest{0.993} & 30.95 & 0.980 & \best{33.29} & \best{0.984} & 23.15 & 0.922 & 34.15 & 0.956 & \sbest{33.19} & \best{0.981} & 26.22 & \sbest{0.962} & \sbest{37.45} & \sbest{0.993} & \best{32.19} & \sbest{0.971}\\
$L=3$ & 38.09 & 0.990 & \sbest{32.78} & \best{0.984} & 32.54 & 0.979 & 22.12 & 0.881 & \best{35.36} & 0.955 & 30.23 & 0.967 & \sbest{27.73} & 0.956 & 35.06 & 0.985 & 31.74 & 0.962\\
$L=4$ & 35.83 & 0.984 & \best{32.93} & 0.982 & 30.39 & 0.969 & 21.06 & 0.855 & 34.38 & 0.947 & 27.74 & 0.946 & \best{28.17} & 0.952 & 32.58 & 0.974 & 30.39 & 0.951\\
$L=5$ & 32.89 & 0.976 & 30.71 & 0.977 & 27.68 & 0.959 & 20.20 & 0.825 & 32.64 & 0.933 & 25.43 & 0.923 & 26.11 & 0.927 & 29.09 & 0.960 & 28.10 & 0.934\\\hline
Linear & 27.77 & 0.973 & 23.10 & 0.921 & 26.68 & 0.959 & 22.27 & 0.922 & 17.39 & 0.869 & 24.98 & 0.946 & 26.82 & 0.955 & 33.98 & \sbest{0.993} & 25.37 & 0.942\\
Quadratic & 29.40 & 0.978 &23.44 & 0.926  &27.51 & 0.963 &22.45 & 0.924 &17.70 & 0.876 &25.70 & 0.950 &26.93 & 0.957 &33.20 & 0.992 &25.79 & 0.946\\
Cubic & 29.98 & 0.979 & 23.71 & 0.928 & 27.76 & 0.964 & 22.37 & 0.921 & 18.04 & 0.884 & 25.96 & 0.951 & 26.25 & 0.954 & 33.51 & 0.992 & 25.95 & 0.947 \\
Spline (5) & 38.87 & \sbest{0.993} & 31.96 & \sbest{0.983} & \sbest{32.96} & \sbest{0.980} & 23.09 & 0.918 & 34.46 & 0.959 & 31.69 & 0.978 & 26.68 & 0.970 & 37.07 & \sbest{0.993} &\sbest{32.10} & \best{0.972}\\
Spline (6) & 38.00 & 0.992 & 31.84 & \best{0.984} & 32.81 & \sbest{0.980} & 22.25 & 0.903 & \sbest{35.24} & \best{0.965} & 30.87 & 0.974 & 26.98 & \best{0.974} & 35.94  & 0.991 & 31.74 & 0.970\\
Linear\,(scale)\!\! & 38.32 & \sbest{0.993} & 30.91 & 0.980 & 32.55 & \best{0.984} & \sbest{23.87} & \sbest{0.930} & 34.43 & \sbest{0.956} & \best{33.43} & \best{0.981} & 25.45 & 0.961 & 36.58 & \sbest{0.993} & 31.94 & \best{0.972}\\\bottomrule
\end{tabular}
}
\scriptsize
\end{table}

\begin{figure*}[tb]
  \centering
  \bgroup 
  \def\arraystretch{0.2} 
  \setlength\tabcolsep{0.2pt}
  \begin{tabular}{ccc}
    Ours without $\loss_{\rm flow}$ & & Ours \\\\
    \includegraphics[width=0.36\linewidth]{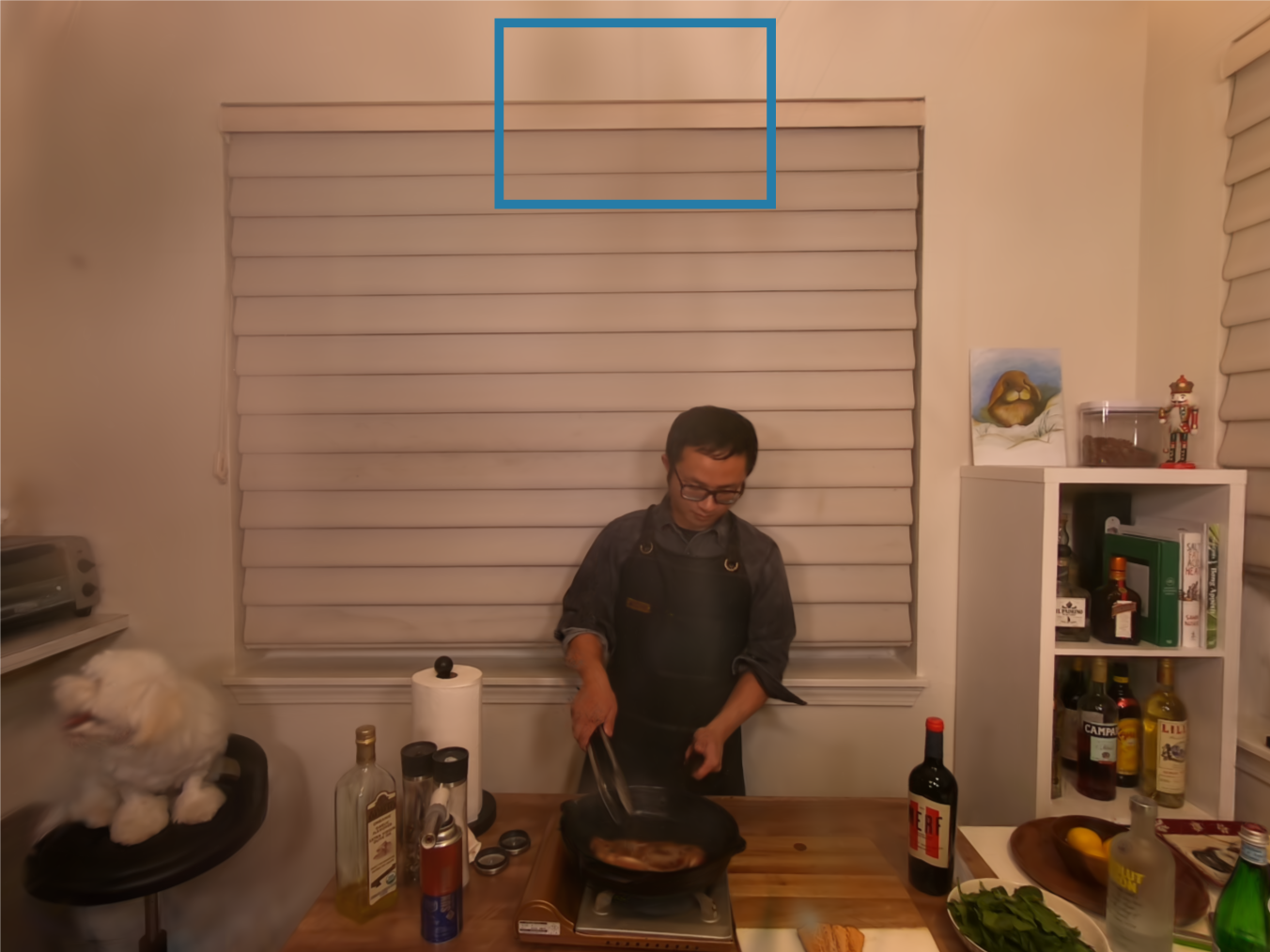} &
     \includegraphics[width=0.1992\linewidth]{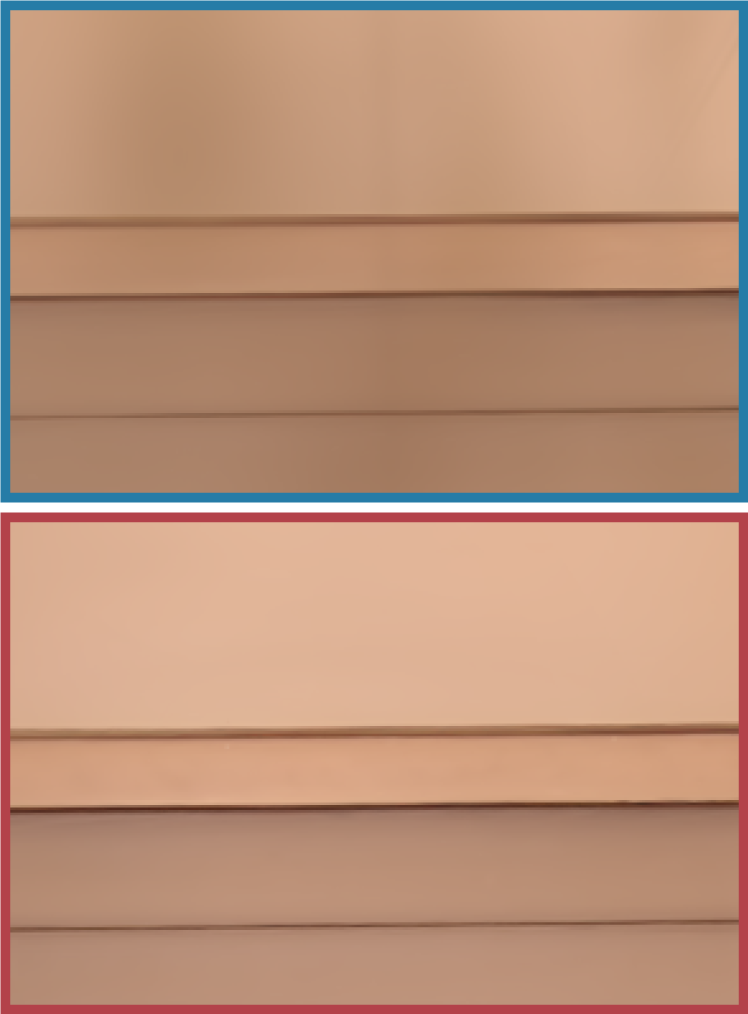} &
    \includegraphics[width=0.36\linewidth]{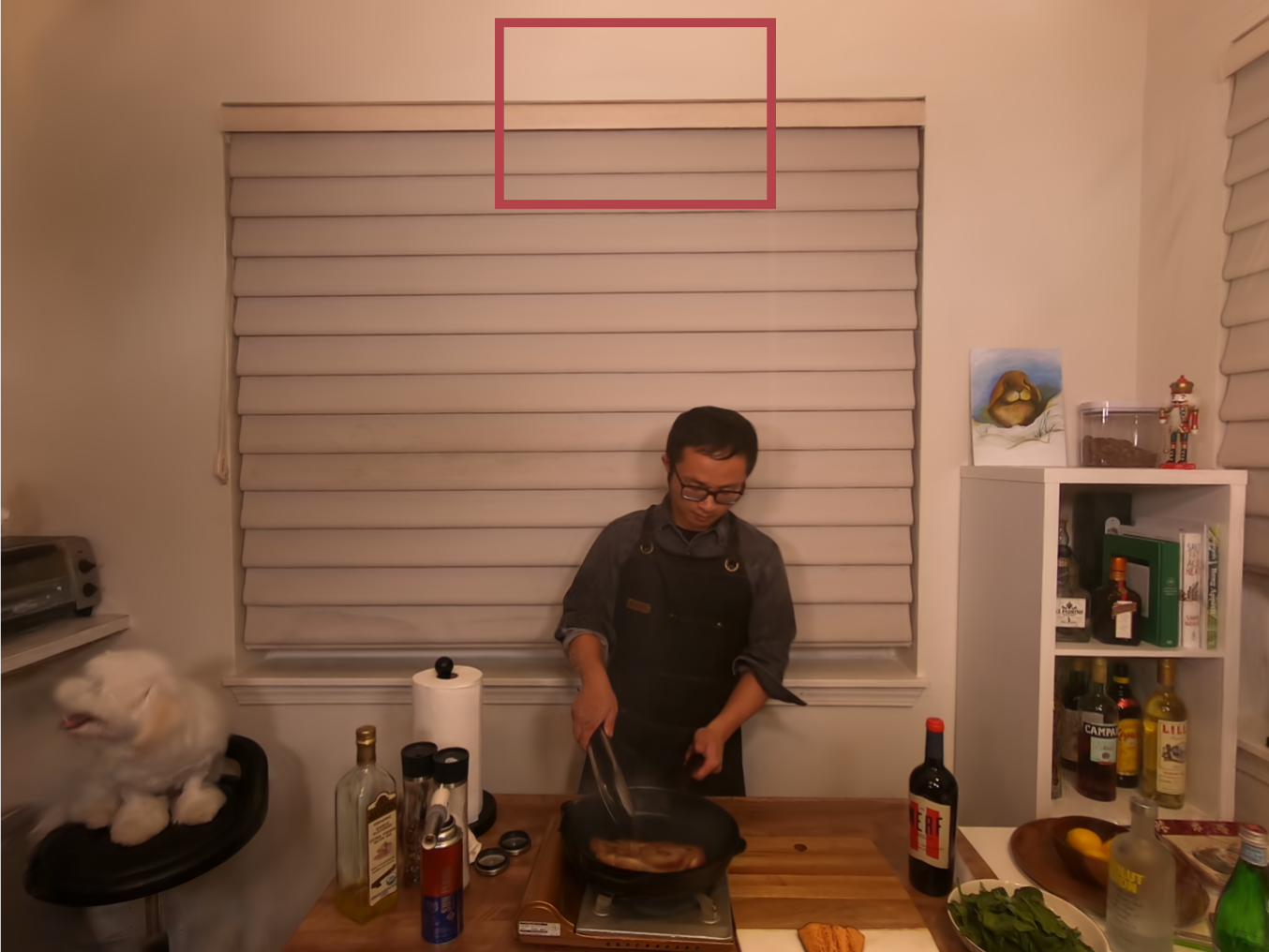} \\

  \end{tabular}\egroup
\caption{Qualitative comparison of disabled and enabled flow loss on DyNeRF. We highlight the difference by zoom view.}\label{fig:flow_loss}
\end{figure*}

\begin{figure}[t]
  \centering
  \bgroup 
  \def\arraystretch{0.2} 
  \setlength\tabcolsep{0.2pt}
  \begin{tabular}{cc}
  \begin{tikzpicture}
    \node[anchor=south west,inner sep=0] (image) at (0,0) {\includegraphics[width=0.40\linewidth,trim={100px 0px 100px 100px},clip]{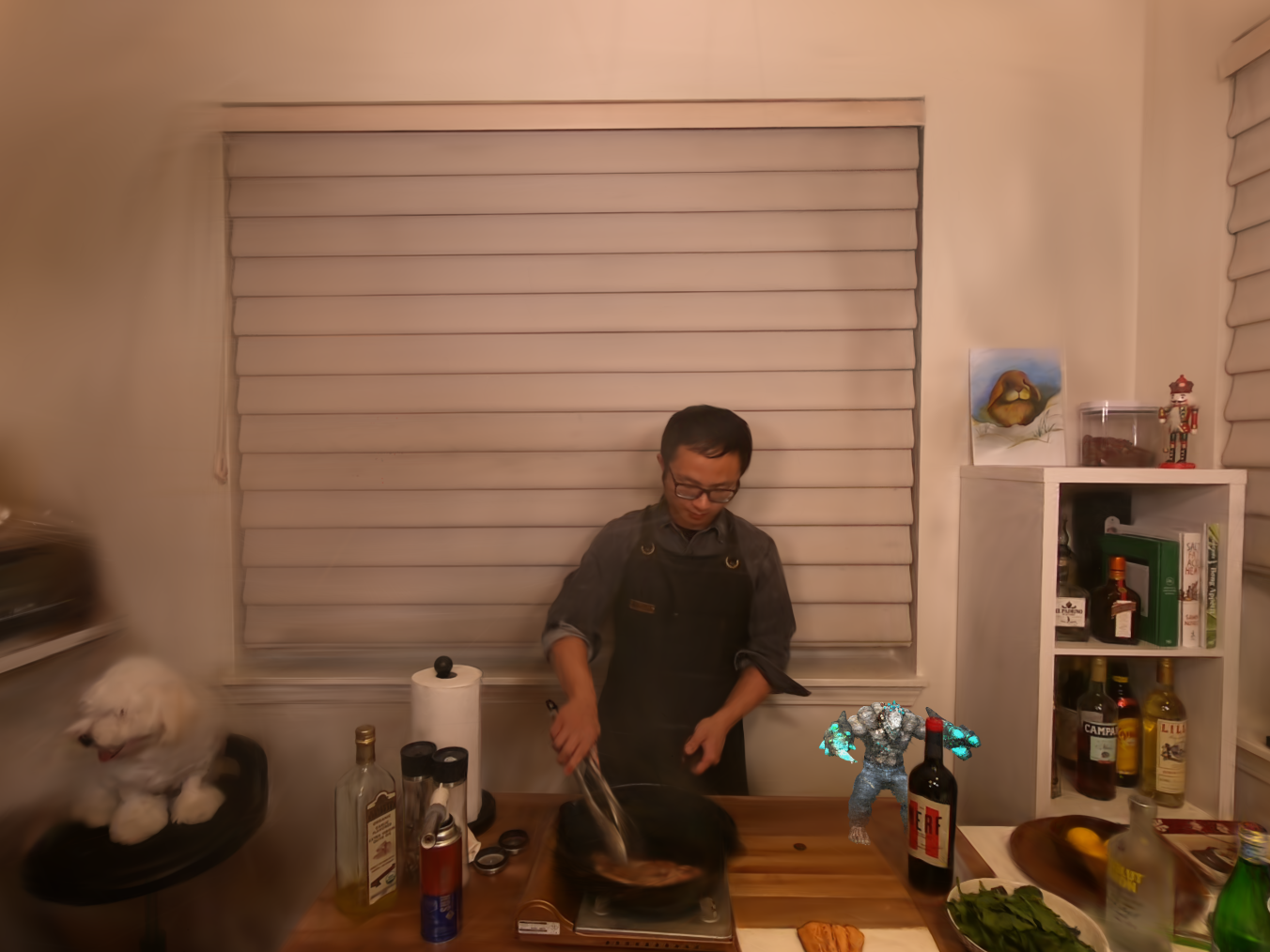}};
    \begin{scope}[x={(image.south east)},y={(image.north west)}]
        \draw[purple, thick] (0.64,0.1) rectangle (0.86,0.32);
    \end{scope}
\end{tikzpicture} & 
  \begin{tikzpicture} 
    \node[anchor=south west,inner sep=0] (image) at (0,0) {    \includegraphics[width=0.40\linewidth,trim={100px 0px 100px 100px},clip]{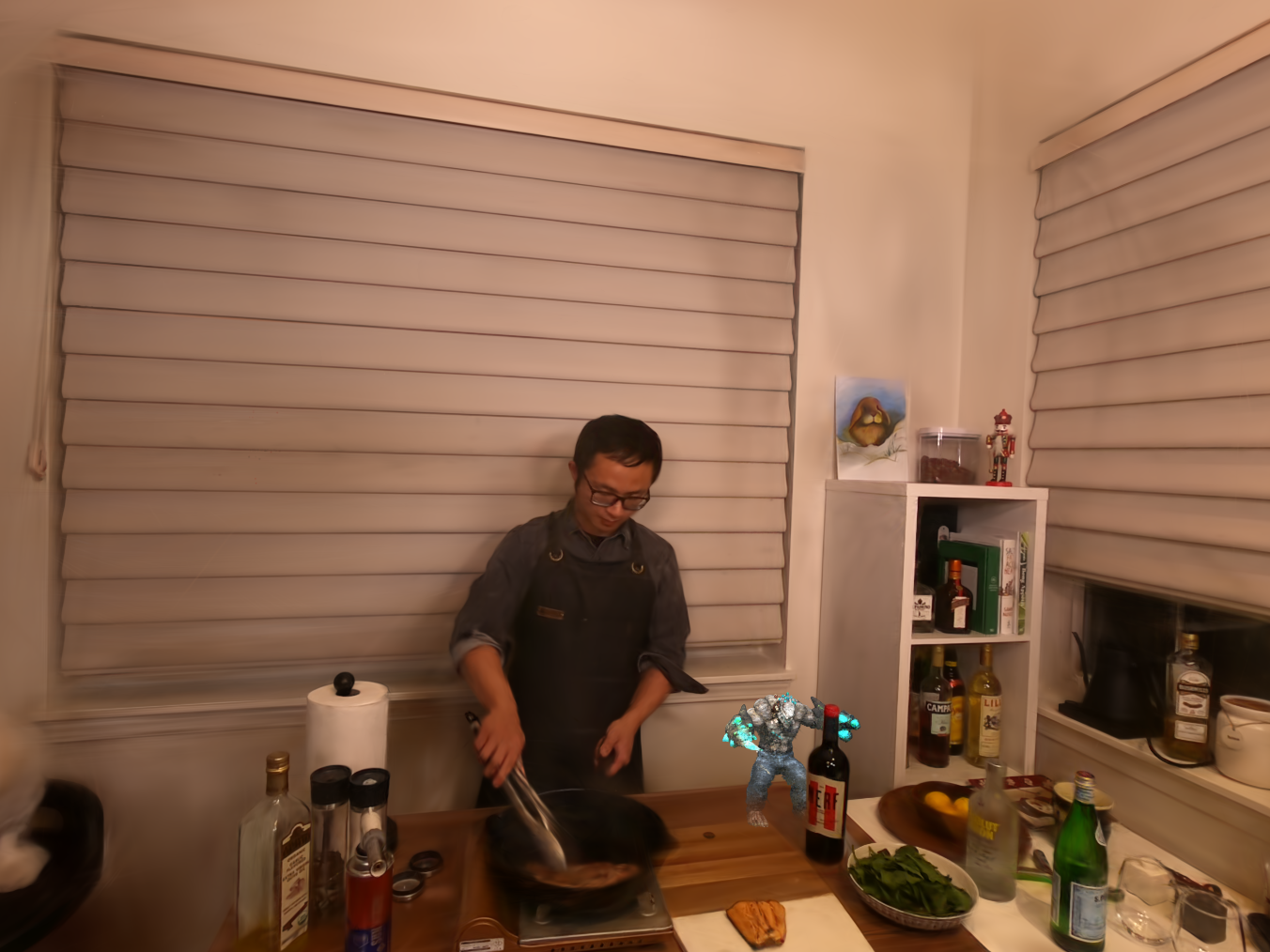}};
    \begin{scope}[x={(image.south east)},y={(image.north west)}]
        \draw[purple, thick] (0.55,0.10) rectangle (0.75,0.34);
    \end{scope}
\end{tikzpicture} \\
  \end{tabular}\egroup
\caption{Composition of two scenes. Our method allows adding 3D objects represented 3D Gaussian into a 3D Gaussian scene. We highlight the added object.}\label{fig:editing}
\end{figure}
\subsection{Experimental results}

\noindent \textbf{Quantitative results.} 
The quantitative results on the D-NeRF dataset are detailed in \cref{tb:dnerf}.
Our method demonstrates performance comparable to TiNeuVox and K-Planes in terms of visual quality indicated by PSNR, SSIM, and LPIPS. Notably, it excels in training time, FPS, and memory size, boasting a rendering speed that is $300\times$ faster than K-Planes.
Furthermore, our method surpasses both 3DGS and D-3DGS in terms of visual quality without compromising rendering speed.
In the DyNeRF scenes experiment, detailed in \cref{tb:dynerf}, while our method does not exceed the baseline in reconstruction quality, it shows a substantial improvement in FPS. Since the DyNeRF scenes contain multi-view data, the D-3DGS baseline naturally improves static 3DGS, unlike monocular scenes. Impressively, our method even attains rendering speeds that exceed real-time performance at a high resolution of 1,354$\times$1,014.
In the challenging HyperNeRF dataset, which is captured by only two moving cameras, referenced in \cref{tb:hypernerf}, our method not only demonstrates rapid rendering speeds but also attains the average PSNR and SSIM scores over the compared methods.

\noindent \textbf{Qualitative results.} 
\Cref{fig:dnerf,fig:dynerf,fig:hypernerf} 
show that our method yields faithful reconstruction for the dynamic scenes.
Unlike the structured representation, which has a fixed size of grids, the unstructured nature of 3D Gaussians enables the control of the expressiveness of the representation adaptively, delivering sharper renderings.

\noindent
\textbf{Effect of the number of parameters $L$.}
\Cref{tb:ablation_l} shows per-scene PSNR and SSIM scores of K-Planes and our method with the different $L$ (\cref{eq:center}).
It is observed that the optimal $L$ for novel view synthesis varies from scene to scene, highlighting the necessity for complex approximations to capture intricate motions effectively.

\noindent
\textbf{Effect of flow loss.}
Additionally, visual comparisons drawn from our method without and with the flow loss (\cref{fig:flow_loss}) reveal that incorporating the flow loss mitigates ghostly artifacts and significantly enhances the accuracy of color reconstruction. 

\noindent
\textbf{Design choice.}
Our method is very flexible and allows arbitrary approximation functions and the choice of time-varying parameters. 
  \Cref{tb:ablation_l} also shows the experimental results of other options for the design of the model to facilitate future dynamic scene reconstruction.
The Linear, Quadratic, and Cubic baselines approximate time-varying 3D positions with polynomials of degree one, two, and three, respectively. The Spline (5) and Spline (6) baselines approximate 3D positions with spline approximations of five and six points, respectively. The Linear (scale) baseline approximates time-varying scales with the linear approximation in addition to positions and rotations. Although a Spline baseline gives minor performance gains in some cases, it achieves 91 FPS for rendering, showing slower rendering than the proposed method. 
The Linear (scale) baseline does not show additional parameters worth performance gains. 
For faster rendering and compact representation, we use the Fourier approximation for 3D positions and model 3D positions and rotations as time-varying parameters.

\noindent
\textbf{Scene composition.}
Since our dynamic 3D Gaussian representation still uses pure 3D Gaussian representation, the learned representation is easy to edit Gaussians. We demonstrate the composition of two scenes with our representation. 
\Cref{fig:editing} illustrates this by combining the \textsc{Mutant} scene from the D-NeRF dataset with the \textsc{Seared Steak} scene from the DyNeRF dataset. 
This demonstrates our method's capability in editing dynamic 3D scenes. 

\section{Discussion and Conclusion}
\noindent
\textbf{Limitations and future directions.}
Our dynamic Gaussians are defined through all times 
of the dynamic scene. This representation implicitly assumes
Gaussians exist over all times of the scene. It enables us to model
naturally the rigid and non-rigid deformation in the scene. On the
other hand, modeling the change of topology, the occurrence of
Gaussians, and extinction of Gaussians (\eg, fluid) is tough. 
The reconstruction capability of the method depends on the number of parameters, so that the scene representation is compact but results in poor rendering quality for very long sequences, requiring additional memory consumption for proper reconstruction.
To overcome these limitations, considering the lifetime of Gaussians, such as adding
start and end time parameters, will model changes in scene topology, and the adaptive decision of flexibility leads to better trade-offs between quality and memory size.

Our Gaussian representation sacrifices the continuity and smoothness of neural field-based volume rendering, and 3DGS causes performance drops in inaccurate camera poses and degradation of generalization performance. 
Distilling NeRFs into our proposed representation in a manner similar to PlenOctree is a potential extension of our method, promising to enhance rendering quality without compromising our advantage in fast rendering.

\noindent
\textbf{Conclusion.}
We present a compact dynamic 3D Gaussian representation that allows the faithful reconstruction of dynamic scenes and real-time rendering.
We propose a representation for the position and rotation of 3D Gaussians as a function of time for modeling the motion of the scene.
The parameterized functions of time introduce memory efficiency and robustness to the number of views per timestep.
Furthermore, we introduce the flow loss that constrains the scene flow of learned Gaussian representation with the ground truth flow.
Our experiments on synthetic and real datasets show that the proposed method achieves real-time dynamic scene rendering
even for a high resolution. %

\bibliographystyle{splncs04}
\bibliography{main}

\end{document}